%% file: main.tex
\documentclass[11pt,a4paper]{article}
\usepackage[margin=19mm]{geometry}
\usepackage[T1]{fontenc}
\usepackage{lmodern}
\usepackage{amsmath,amssymb,booktabs,graphicx,textcomp,longtable,array,float}
\usepackage[numbers,sort&compress]{natbib}
\usepackage[font=small,labelfont=bf]{caption}
\usepackage{microtype}
\usepackage{tabularx,xcolor,fancyhdr,placeins}
\usepackage[hidelinks]{hyperref}
\title{\bfseries The limits of reconstruction fidelity and evaluation in medical image synthesis with latent diffusion}
\author{Marvin Seyfarth\textsuperscript{1,2,3,*},
Yannik Frisch\textsuperscript{1,2,3},\\
Sandy Engelhardt\textsuperscript{1,2,3}, and
Salman Ul Hassan Dar\textsuperscript{1,2,3}}
\date{}
\hypersetup{
  pdftitle={The limits of reconstruction fidelity and evaluation in medical image synthesis with latent diffusion},
  pdfauthor={Marvin Seyfarth, Yannik Frisch, Sandy Engelhardt, Salman Ul Hassan Dar}
}

\begin{document}
\maketitle

\begin{center}
\small
\textsuperscript{1}Institute for Artificial Intelligence in Cardiovascular Medicine, Medical Faculty of Heidelberg University, Heidelberg University, Germany\par
\textsuperscript{2}Department of Cardiology, Angiology, Pneumology, Heidelberg University Hospital, Heidelberg, Germany\par
\textsuperscript{3}DZHK (German Centre for Cardiovascular Research), Partner Site Heidelberg/Mannheim, Heidelberg, Germany\par
\smallskip
\textsuperscript{*}Correspondence to \href{mailto:marvin.seyfarth@med.uni-heidelberg.de}{\nolinkurl{marvin.seyfarth@med.uni-heidelberg.de}}
\end{center}

\begin{abstract}
Latent diffusion models (LDMs) enable efficient, high-quality medical image synthesis by learning data distributions in an autoencoder's latent space. In practice, this autoencoder is selected for reconstruction fidelity and frozen before generator training. Whether this yields the best generator is rarely tested, and generative evaluation is uncertain, as common metrics depend on pretrained feature extractors and capture only specific properties. We address both in 3D lung CT and 2D chest radiography. First, across 233 controlled failure settings, no single metric detected every failure, detection depended on the feature extractor, and combining metrics improved severity prediction. Second, across 27 CT and 24 radiographic LDM pipelines, the autoencoder with the lowest reconstruction error did not yield the lowest Fréchet Inception Distance, the most widely used generative metric, and changing generator capacity or architecture did not consistently resolve this mismatch. Autoencoders should therefore be jointly evaluated with their generators using complementary, representation-aware metrics.
\end{abstract}

\section*{Highlights}
\begin{itemize}
\item Controlled metrics benchmark reveals: No generative evaluation metric for medical image synthesis detects every tested failure, and sensitivity depends on feature representation.
\item Reconstruction metrics show strong correlations with one another, whereas generative metrics rank pipelines differently.
\item Complementary metrics improve severity prediction beyond Fréchet inception distance (FID) in selected exploratory comparisons.
\item Preferred autoencoder compression depends on the downstream generator and evaluation metric.
\end{itemize}

\noindent\textbf{Keywords}\enspace
latent diffusion, VQ-GAN, medical imaging, reconstruction fidelity, generative evaluation.

\begin{figure}[t]
\centering
\includegraphics[width=\textwidth]{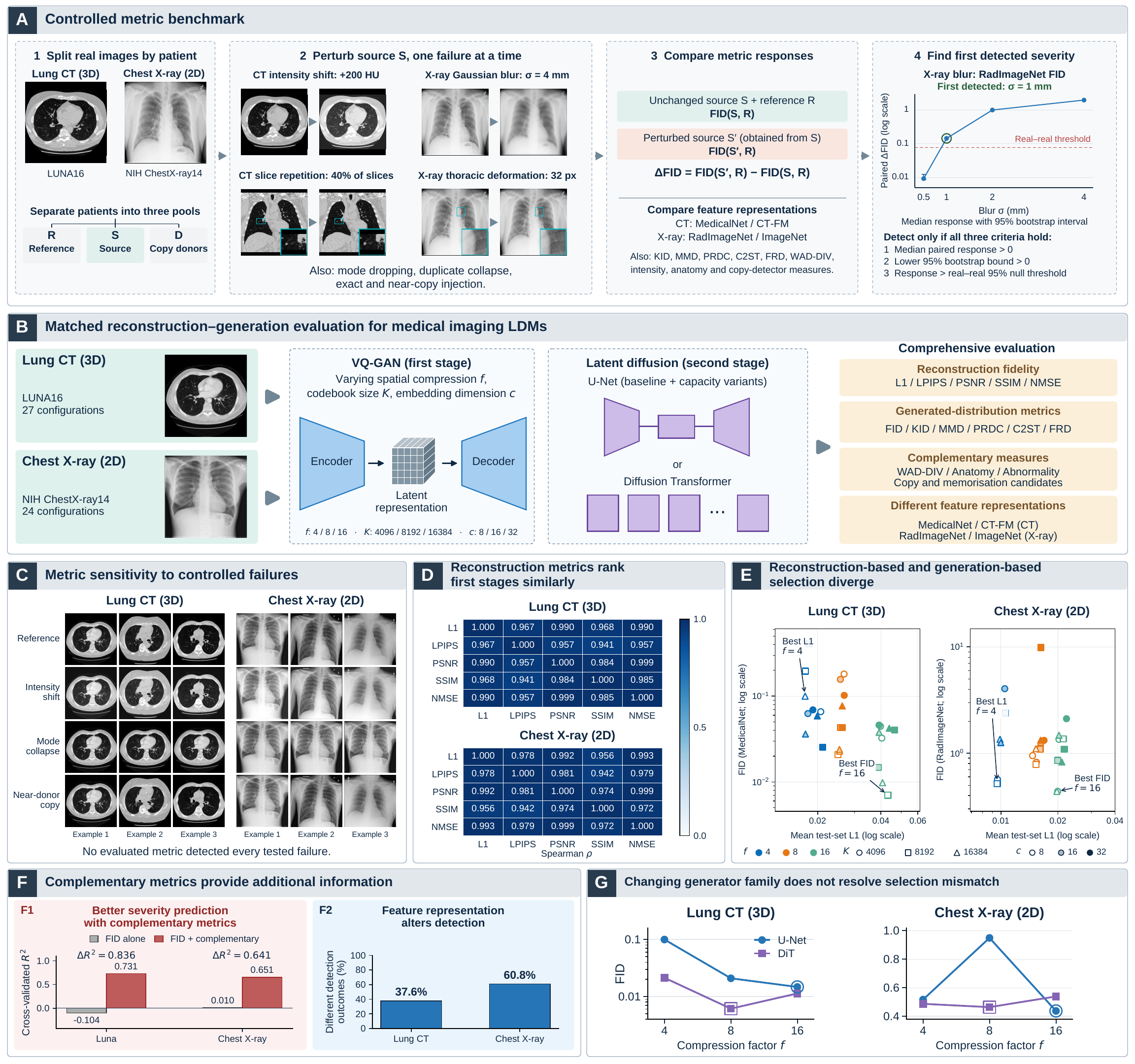}
\caption{Study overview.
(A) Metric benchmark using patient-disjoint cohorts, controlled image failures, and comparisons across feature representations. The curve shows RadImageNet FID under chest X-ray blur.
(B) Matched evaluation of autoencoder reconstruction and LDM generation.
(C) Examples of controlled image failures.
(D) Rank correlations between reconstruction metrics after aligning their directions.
(E) Reconstruction-based versus FID-based autoencoder selection across model configurations.
(F) Complementary severity prediction and feature-dependent failure detection. F1 illustrates CT mode dropping (MedicalNet zero-reference WAD-DIV) and X-ray deformation (ImageNet C2ST); F2 counts matched endpoint--failure comparisons.
(G) U-Net versus DiT FID.}
\label{fig:study-overview}
\end{figure}

\section{Introduction}
\label{sec:introduction}

Recent advances in generative modeling have substantially improved the ability to synthesize realistic medical images, offering opportunities to augment training datasets and improve coverage of underrepresented groups \citep{gao2023syntheticxray,ktena2024generativefairness}. Synthetic images may also support data sharing, which requires assessing the risk of reproducing training images \citep{dar2026memorization}.
Latent diffusion models (LDMs) have become a prominent backbone for medical image generation. They follow a two-stage design in which an autoencoder first maps images to a compressed latent representation and a diffusion model subsequently learns the data distribution within that latent space \citep{rombach2022ldm,pinaya2022brainldm,wang2025_3d_meddiffusion,guo2025maisi,seyfarth2026medlord,khader2023meddiffusion}. Working in this smaller space reduces the cost of generation, particularly for high-resolution volumes. In a typical workflow, candidate autoencoders are compared using reconstruction metrics. One configuration is selected and frozen, and a downstream diffusion model is trained in its latent representation \citep{mullerfranzes2023medfusion}. This choice fixes the spatial resolution of the latent grid and determines which image information remains available to the generator \citep{rombach2022ldm,chen2025dcae}. However, studies outside the medical domain suggest that reconstruction fidelity and downstream generative performance do not necessarily improve in parallel \citep{yao2025reconstruction,skorokhodov2025diffusability}. This raises a central question for medical image synthesis: does selecting an autoencoder for reconstruction quality also identify the configuration preferred after downstream generative training?
\citet{khader2023meddiffusion} compared vector-quantised autoencoder (VQ-GAN) compression factors of 8 and 4 and adopted the latter after evaluating reconstruction quality and radiologist-assessed anatomical retention. This provides a clinically motivated basis for selection. However, such studies usually do not train a matched diffusion model for each candidate autoencoder and compare their generated images. It therefore does not establish whether the reconstruction-based preference would persist after generative training. Answering that question requires evaluating each candidate autoencoder together with its downstream generator and a decision which properties of generated images to measure. Generated medical images should preserve plausible anatomy, clinically relevant features and the diversity found in real medical data \cite{Tudosiu2024}.
Many widely used quantitative measures originate in general computer vision and compare real and generated cohorts through representations extracted by pretrained neural networks \cite{heusel2017ttur,binkowski2018mmd,naeem2020prdc}.
In medical imaging, the interpretation of such comparisons could depend on the selected feature extractor, its pretraining domain and cohort, the represented anatomy and pathology, the reference cohort, preprocessing, the unit of observation, and the available sample support. \\
These limitations of metrics have motivated multifaceted evaluation approaches that explicitly define the property assessed by each metric and the scope of its interpretation \cite{woodland2024features,deo2025metrics,zamzmi2025scorecard}. As a consequence, throughout this study, a \emph{metric} is always defined together with the feature representation, reference cohort, and evaluation role in which it is applied.
Measures of distributional agreement \cite{konz2026frd}, local diversity \cite{naeem2020prdc,seyfarth2026waddiv}, training proximity \cite{dar2026memorization} and anatomy \cite{wasserthal2023totalsegmentator} capture different properties. Agreement among complementary measures can strengthen an interpretation, while averaging them can conceal trade-offs. The scorecard of \citet{zamzmi2025scorecard} places quantitative measures alongside task-based and human evaluation, ethics and documentation of the reference data. The choice of measures depends on the intended use of the images \citep{xing2023dontperfectamazingunveil}.
\citet{deo2025metrics} tested metrics using known failures in brain MRI and MR angiography, including noise, local tumour alterations, duplication, mode collapse, scanner-density shifts and mode invention. They also compared variational autoencoders, adversarial generators and diffusion models. Some distributional metrics missed local anatomical alterations or repeated images. Others yielded more favourable scores when training images were copied into the evaluated cohort or when rare anatomical groups were removed. Generator rankings also differed from those based on estimated tumour and vessel segmentation quality. These findings show how the choice of metric can affect which shortcomings are apparent in generated images. Our study connects this evaluation problem to autoencoder selection. The benchmark informs how we interpret metric agreement and disagreement in the matched pipeline study.
For LDMs, this has implications beyond evaluating the final synthetic images: the choice of metrics can also influence which autoencoder, and thus which latent representation, is preferred for downstream generation.

We address these research gaps through two experiments (Fig.~\ref{fig:study-overview}A,B). First, a controlled generative metrics benchmark tests which applied failures are detected and how detection depends on the feature representation used for evaluation. Second, a matched VQ-GAN--LDM study tests whether selecting autoencoders by reconstruction quality yields the same preference as evaluating their downstream generated images. We conduct both experiments in 3-D lung CT and 2-D chest radiography. These modalities allow us to assess whether findings persist across differences in dimensionality and image variability. Both are established applications of medical image synthesis \cite{oliveras2026land,mullerfranzes2023medfusion}. The matched study comprises 27 lung CT and 24 chest-X-ray first-stage configurations, varying spatial compression, codebook size, and embedding dimension, each evaluated through a common dataset-specific U-Net training protocol. Additional analyses vary U-Net capacity and replace the U-Net with a diffusion transformer to examine whether first-stage preferences depend on the downstream architecture \cite{seyfarth2026volditcontrollablevolumetricmedical}. Finally, HA-GAN \cite{hagan2022} and wavelet-diffusion models \cite{phung2023wavediff,friedrich2024wdm} provide contextual comparisons between complete generative pipelines.

This systematic empirical study makes four contributions:
\begin{enumerate}
\item \textbf{Testing metric sensitivity to known failures:} A controlled benchmark across lung CT and chest radiography tests how commonly used generative metrics respond to predefined intensity, image-quality, mode, diversity, spatial, anatomical, continuity, and copy-injection failures. Paired replicates, clean real--real comparisons, and predefined detection rules establish which failures each metric detects (Section~\ref{sec:results_controlled_metric_benchmark}; Fig.~\ref{fig:study-overview}A,C).
\item \textbf{Testing metric dependence on feature representation:} Matched comparisons across evaluation feature extractors determine whether changing the representation alters failure detection or the first detected severity, clarifying how strongly metric interpretation depends on the features used (Section~\ref{sec:results_controlled_metric_benchmark}; Fig.~\ref{fig:study-overview}F2).
\item \textbf{Testing reconstruction-based autoencoder selection:} A controlled study of 27 lung CT and 24 chest-X-ray VQ-GAN--LDM pipelines compares autoencoder selection by reconstruction quality with selection by the generated-distribution measure, Fréchet inception distance (FID). Every autoencoder is evaluated through a common dataset-specific downstream training protocol (Section~\ref{sec:results_selection_mismatch}; Fig.~\ref{fig:study-overview}D,E).
\item \textbf{Informing model selection with complementary evidence:} Complementary metrics and codebook diagnostics reveal trade-offs in autoencoder--generator selection. Matched U-Net capacity and diffusion-transformer comparisons assess whether model preferences persist across downstream architectures (Fig.~\ref{fig:study-overview}G), while contextual HA-GAN and wavelet-diffusion comparisons extend the analysis to complete generative pipelines (Sections~\ref{sec:results_selection_mismatch}--\ref{sec:results_generator_family}).
\end{enumerate}

\section{Methods}
\label{sec:methods}

\subsection{Study design, datasets and preprocessing}
\label{sec:methods_study_design}

We conducted two complementary experiments in 3-D lung CT and 2-D chest radiography (Fig.~\ref{fig:study-overview}A,B). First, we introduced known failures into real images and measured whether these failures can be detected by changes in the evaluation metrics. Second, we trained matched autoencoder-diffusion pipelines and compared preferences based on reconstruction quality with those based on generated images. Additional analyses examined complementary metrics, codebook usage and the effect of changing the downstream generator.
The CT study used 888 LUNA16 scans, derived from LIDC-IDRI, with 800 training, 39 validation and 49 test volumes \citep{SETIO20171}. The radiography study used a fixed subset of 20,000 NIH ChestX-ray14 images \citep{wang2017chestxray8}, divided into 10,000 training, 5,000 validation and 5,000 test images. Splits were random at the patient level.
CT intensities were clipped to $[-1000,1000]$ Hounsfield units (HU), and volumes were centre-cropped or padded to $512\times512\times256$ voxels. Chest radiographs were min-max normalised separately for each image. Generated CT volumes were restored to HU. Preprocessing for individual evaluation feature extractors is described in Supplementary Table~\ref{tab:supp-extractor-preprocessing}.

\subsection{Evaluation framework}
\label{sec:methods_evaluation_framework}

\paragraph{Reconstruction and generation.}

Reconstruction was assessed by comparing each decoded image with its original input. Mean test-set L1, the mean absolute reconstruction error, was the predefined primary reconstruction metric. Perceptual loss (LPIPS), peak signal-to-noise ratio (PSNR), structural similarity (SSIM) and normalised mean squared error (NMSE) provided complementary measures of the same paired task.
We assessed generation by comparing cohorts of generated and real images. FID was the primary generative metric. Additional measures were kernel inception distance (KID), Fréchet radiomics distance (FRD), maximum mean discrepancy (MMD), precision, recall, density, coverage and Wasserstein distance diversity (WAD-DIV). The classifier two-sample test (C2ST) measures how readily the classifier distinguishes real and generated cohorts from their image features. We used nearest-centroid classification with stratified five-fold cross-validation and report its absolute balanced-accuracy gap from chance, $|\mathrm{BA}-0.5|$.
The WAD-DIV variants were interpreted separately. The empirical-reference version measures disagreement between local-distance distributions in real and generated features, so smaller raw values indicate closer agreement. Its normalised agreement form has the opposite direction: larger values indicate closer agreement. The zero-reference version measures spread and was interpreted relative to the corresponding clean real-cohort value. Supplementary Table~\ref{tab:metric-overview} provides an overview of the metrics, their preferred directions and the implementation settings.

\paragraph{Evaluation features and reference images.}

Our primary evaluation features came from ResNet-50 encoders pretrained on medical images: MedicalNet for CT and RadImageNet for chest radiography \citep{chen2019med3d,doi:10.1148/ryai.210315}. CT-FM and ImageNet-pretrained Inception-v3 provided alternative representations for CT and radiography, respectively \citep{projectlighter_ctfm,Szegedy2015RethinkingTI}. We use the dataset names to identify the evaluation encoders throughout the study.
For the primary comparison of trained pipelines, CT metrics compared all 49 held-out test volumes with 800 generated volumes per configuration. Radiography metrics compared 5,000 real test images with 5,000 generated images. These reference cohorts were separate from the training-derived pools used in the controlled benchmark described below.

\paragraph{Proximity to training data.}

Training data memorisation was assessed using dataset-specific learned detector embeddings and two directional counts \citep{dar2026memorization}. $N_{\mathrm{copy}}$ counted generated images whose maximum Pearson correlation with any training image exceeded a fixed threshold, thereby characterising the number of synthetic samples that were patient data copies. $N_{\mathrm{mem}}$ counted training images whose maximum correlation with any generated image exceeded the same threshold, which are the patient data samples classified as memorised. The first count therefore describes generated candidates, and the second describes the training cases they resemble.
For each dataset, the threshold was calibrated at the 95th percentile of training-to-nearest-validation correlations and then held fixed across configurations. Detector implementations and threshold calibration are described in Supplementary Section~\ref{sec:supp-detector-methods}. Results appear in Supplementary Tables~\ref{tab:supp-grid-Luna}, \ref{tab:supp-grid-ChestXray}, \ref{tab:supp-capacity-results} and~\ref{tab:supp-generators}.

\subsection{Controlled metrics benchmark}
\label{sec:methods_controlled_metric_benchmark}

\paragraph{Creating known failures.}

We tested metric sensitivity by making controlled changes to real images (Fig.~\ref{fig:study-overview}A,C). These changes served as surrogates for failures that can occur in generated data. Training cases were partitioned once into three patient-disjoint pools: real reference images $R$, source images $S$ and donor images $D$ for copy-injection experiments. Each fixed-seed replicate $r$ selected subsets $R_r$, $S_r$ and, for copy-injection analysis $D_r$ from those pools.
For image-level changes, the perturbed cohort $P_{j,s,r}$ contained the same source cases as $S_r$, with failure subtype $j$ applied at severity $s$. Other experiments changed the selection or frequency of images within the cohort. These included mode dropping, which removes part of the represented variation; diversity collapse, which reduces the variety of cases; and injection of exact or near-copies from the donor pool. Each failure subtype and severity was generated separately. Failures were not combined.
Altogether the benchmark covered intensity shifts, degradation of image quality, mode dropping, diversity collapse, discontinuities across CT slices, thoracic or organ-specific deformation, and exact or near-copy injection. Each failure had four non-zero severity levels, $s_1$ to $s_4$, ordered from least to most severe. Every level was generated directly from the clean source cohort, rather than by adding changes to the previous level. Supplementary Section~\ref{sec:supp-benchmark-methods} gives pool sizes, random seeds and severity coding. Exact transformations and severity levels are listed in Supplementary Tables~\ref{tab:supp-benchmark-transformations} and~\ref{tab:supp-benchmark-transformations-chest}; examples are shown in Supplementary Fig.~\ref{fig:supp-perturbation-examples}.

\paragraph{Measuring responses and defining detection.}

For each metric, we evaluated both the perturbed cohort $P_{j,s,r}$ and its unchanged source cohort $S_r$ against the same reference subset $R_r$ selected for replicate $r$. We oriented the responses so that positive values indicated a change in the expected failure direction.
Let $E_m(A;\mathcal{R}_{m,r})$ be the raw value of metric $m$ for cohort $A$, using its metric-specific reference $\mathcal{R}_{m,r}$. With orientation function $g_{m,r}$, the paired response was:
\begin{equation}
\Delta_{m,j,s,r}=g_{m,r}\!\left[E_m(P_{j,s,r};\mathcal{R}_{m,r})\right]-g_{m,r}\!\left[E_m(S_r;\mathcal{R}_{m,r})\right].
\label{eq:controlled_metric_response}
\end{equation}
The orientation function was specified for each metric: FID, KID, MMD, the C2ST gap and raw empirical-reference WAD-DIV were left unchanged $g(E)=E$; precision, recall, coverage and normalised empirical-reference WAD-DIV agreement were complemented $g(E)=1-E$, and density was sign-reversed $g(E)=-E$. For zero-reference WAD-DIV, it used the absolute deviation from the replicate's clean anchor. A response in the opposite direction remained negative and was not counted as a detection.
As an example for FID, which increases as the measured distributions separate, the response was simply:

\begin{equation}
\Delta_{\mathrm{FID},j,s,r}=\mathrm{FID}(P_{j,s,r},R_r)-\mathrm{FID}(S_r,R_r).
\label{eq:controlled_metric_response_fid}
\end{equation}
For each metric, failure subtype and severity, we summarised the paired responses by their median across replicates. We obtained a 95\% benchmark-bootstrap interval by resampling the replicate-level paired responses with replacement 2,000 times and recalculating the median after each resampling:

\begin{equation}
\widetilde{\Delta}_{m,j,s}=\operatorname{median}_{r}\Delta_{m,j,s,r}.
\label{eq:controlled_metric_median_response}
\end{equation}

We also required the response to exceed a threshold derived from clean real--real comparisons. If $Q_{m,r}$ is the oriented clean real--real value in replicate $r$, this threshold was the 95th percentile of its absolute deviation from the median:

\begin{equation}
T^{\mathrm{RR}}_m=\operatorname{quantile}_{0.95,r}\!\left(\left|Q_{m,r}-\operatorname{median}_{r\prime}Q_{m,r\prime}\right|\right).
\label{eq:controlled_metric_real_real_threshold}
\end{equation}

A non-zero severity was counted as detected only if (1) the median paired response in Eq.~\eqref{eq:controlled_metric_median_response} was positive, (2) the lower bound of its 95\% bootstrap interval, obtained by resampling the replicate-level responses in Eq.~\eqref{eq:controlled_metric_response}, exceeded zero, and (3) the median paired response exceeded the clean real--real threshold in Eq.~\eqref{eq:controlled_metric_real_real_threshold}. The first detected severity was the lowest tested non-zero level satisfying these conditions.

\paragraph{Replicates and cohort sizes.}

The CT benchmark used 20 fixed-seed replicates, each comparing 200 unchanged source volumes with 200 perturbed volumes. For the more computationally expensive CT medical-proxy analyses using CT-CLIP and TotalSegmentator-based anatomical measures (see Section~\ref{sec:methods_primary_analysis}, ``Complementary metrics''), we used 10 replicates, each comparing 100 unchanged and 100 perturbed volumes. Radiography metrics used 20 replicates, each comparing 1,024 unchanged images with 1,024 perturbed images.
We expected Wasserstein distances between these distributions to be less sensitive to limited sample sizes than distances in high-dimensional image-feature spaces \citep{Fournier2015}.

\paragraph{Prediction of failure severity.}

We additionally asked whether a second metric helped predict the severity of a known failure beyond FID alone. For each dataset and failure family, we compared ridge regressions with penalty $\alpha=1$ using either FID alone or FID plus one additional metric. Predictors were oriented metric values, and the target was severity rank divided by four, on a scale from 0 to 1.
We used five-fold cross-validation grouped by benchmark replicate and standardised predictors within each training fold. The examples highlighted in the results were selected after inspection (Fig.~\ref{fig:study-overview}F1).

\subsection{Matched autoencoder--LDM study and selection comparison}
\label{sec:methods_primary_pipeline}

The primary model comparison asked whether the autoencoder with the lowest mean reconstruction L1 also produced the LDM with the lowest held-out FID (Fig.~\ref{fig:study-overview}B,E). We trained a separate diffusion model for every retained autoencoder using a common dataset-specific protocol. This allowed reconstruction quality and generated-image evaluation to be compared across matched pipelines. All retained configurations were trained under the same dataset-specific protocol and evaluated at common checkpoints. No individual configuration or checkpoint was selected on the basis of its evaluation score for this primary grid. Metric-based preferences were determined retrospectively from held-out test metrics and identify the most favourable observed scores within the evaluated grid.

\paragraph{Autoencoder configurations and training.}

Using the VQ-GAN architecture from MedLoRD \citep{seyfarth2026medlord}, we varied spatial compression $f\in\{4,8,16\}$, codebook size $K\in\{4096,8192,16384\}$ and embedding dimension $c\in\{8,16,32\}$. The CT experiment included all 27 configurations in the $3\times3\times3$ grid. In radiography, the three configurations with $f=4$ and $c=32$ did not converge under the common protocol. We retained the remaining 24 and did not retrain the failed configurations with modified settings. Autoencoders were evaluated at fixed dataset-specific checkpoints: epoch 200 for CT and epoch 300 for radiography. Validation losses were monitored, and the fixed checkpoints were chosen so that all retained models had shown convergence behaviour by those epochs. The same checkpoint was used for all retained autoencoders within each dataset to support a fair comparison. Their training objective combined pixelwise, perceptual, quantisation and adversarial losses. Training behavior is shown in Supplementary Fig.~\ref{fig:supp-training-behaviour}.

\paragraph{Diffusion training}

Each trained autoencoder was frozen and used to precompute its training latents. We then trained an unconditional U-Net diffusion model in that latent space, following the MedLoRD configuration \citep{seyfarth2026medlord}. Generated latents were decoded using the same frozen autoencoder. Within each dataset, all configurations received the same number of optimiser updates. Validation losses were monitored, and evaluation used fixed checkpoints at epoch 375 for CT and epoch 400 for radiography. These dataset-specific checkpoints were chosen so that all models had shown convergence behaviour. Dataset-specific settings are listed in Supplementary Table~\ref{tab:supp-training}, with architecture and optimisation details in the accompanying Supplementary Methods.

\subsection{Complementary metrics}
\label{sec:methods_primary_analysis}

We next asked which differences between pipelines were obscured by evaluating generation using FID alone. Generated-image metrics were compared using Spearman rank correlations, separately for each dataset, evaluation feature extractor and reference cohort. We reversed the signs of FID, KID, MMD, the C2ST gap and FRD so that higher oriented values indicated more favourable scores. Precision, recall, density, coverage and normalised empirical-reference WAD-DIV agreement (WAD-A) retained their original directions. We also reversed the signs of $N_{\mathrm{copy}}$ and $N_{\mathrm{mem}}$ so that higher oriented values indicated fewer images classified as copies or as memorised, respectively.
For CT, we used CT-CLIP, a vision-language foundation model, to obtain continuous scores for 18 abnormality concepts \citep{hamamci2024ctclip}. For each concept, we calculated the Wasserstein distance between the score distributions of real and generated volumes. Smaller distances indicated closer agreement and defined a ranking of the 27 configurations.
Anatomical comparisons considered the left lung, right lung, heart and aorta. For each structure, we computed the Wasserstein distance between real and generated organ-volume distributions using only valid masks.
Codebook usage was measured on the training split using the fraction of entries used at least once and the effective code fraction $\exp(H)/K$, where $H$ is the entropy of code assignments and $K$ is the codebook size. We related each diagnostic to primary FID using ordinary-least-squares fits and in-sample $R^2$, separately for each dataset. Further details are provided in Supplementary Section~\ref{sec:supp-architecture}.

\subsection{Sensitivity to downstream architecture and generator family}
\label{sec:methods_architecture_sensitivity}

We tested whether preferences among autoencoders changed with the capacity or architecture of the downstream generator (Fig.~\ref{fig:study-overview}G). In particular, the U-Net comparison examined whether greater denoiser capacity could compensate for differences between latent representations. Within each dataset and compression factor $f\in\{4,8,16\}$, we retained the VQ-GAN with the lowest held-out mean L1. We trained smaller, baseline and wider U-Net diffusion models using the same precomputed latents. Their channel widths were [32, 64, 128, 256], [64, 128, 256, 512] and [96, 192, 384, 768], respectively, corresponding to approximately 0.25, 1 and 2.25 times the baseline parameter count. All variants used the same evaluation procedure as the primary experiment.
We also paired the same selected autoencoders with unconditional diffusion transformers following VolDiT-L4 \citep{seyfarth2026volditcontrollablevolumetricmedical}. Radiography used its 2-D spatial adaptation. These comparisons retained the autoencoder checkpoints, precomputed latents, decoder, data splits, diffusion sampling family and dataset-specific metrics. They tested whether autoencoder preferences persisted when the downstream model changed from a convolutional U-Net to a transformer.
Finally, we evaluated complete generative pipelines without a learned first-stage autoencoder as contextual comparisons. We trained one wavelet-diffusion model and one HA-GAN per dataset: WDM-3D for CT, WaveDiff for radiography and the corresponding dataset-specific HA-GAN variants \citep{friedrich2024wdm,phung2023wavediff,hagan2022}. These systems retained their repository-specific preprocessing, architectures and training protocols. Evaluation used the same dataset-specific held-out reference cohorts as the latent models.

\subsection{Computational resources}

Experiments used high-performance computing nodes with NVIDIA A40, A100 and H200 GPUs and a local workstation with two NVIDIA RTX 4090 GPUs. Training, image generation and evaluation required an estimated 30,000 GPU-hours. %

\section{Results}
\label{sec:results}
We first tested generative metric responses to known failures, then compared matched autoencoder--LDM pipelines and lastly varied the downstream generator (Fig.~\ref{fig:study-overview}A,B).

\subsection{Metric sensitivity depends on the evaluation features and failure}
\label{sec:results_controlled_metric_benchmark}

\begin{figure}[t]
\centering
\includegraphics[width=\textwidth]
    {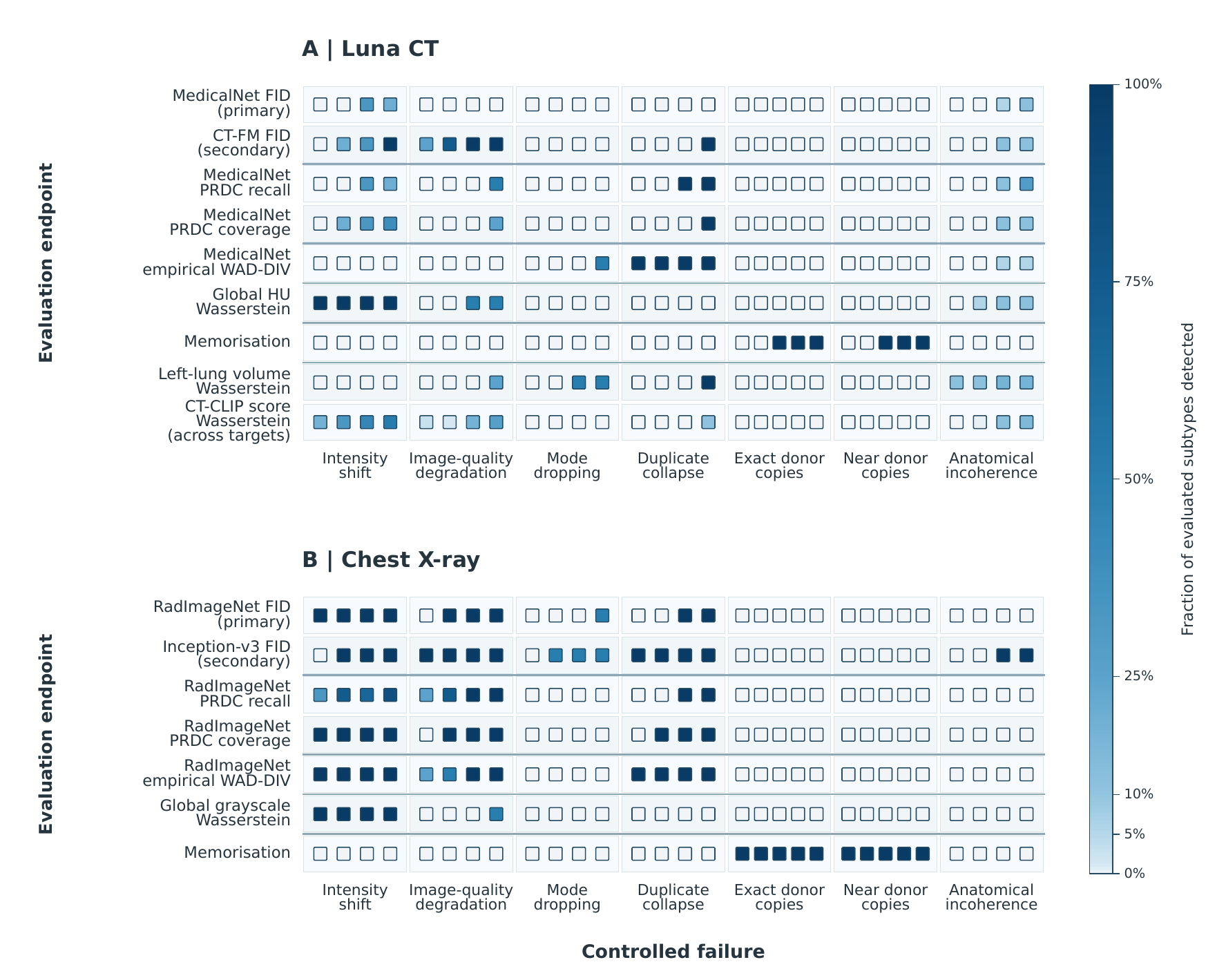}
\caption{Detection of controlled failures in lung CT (A) and chest radiography (B). Each square shows the proportion of failure subtypes detected at a given severity level. Empty squares indicate no detection. Detection required a positive paired response, a bootstrap interval excluding zero, and an effect exceeding clean real--real variation. Severity levels are ordered within each failure type. FID and PRDC use the indicated feature encoders, while HU and grayscale metrics use image intensities. Copy detection refers to controlled donor-copy injection. Exact settings are provided in Supplementary Tables~\ref{tab:supp-benchmark-transformations}--\ref{tab:supp-benchmark-transformations-chest}.}
\label{fig:controlled_failure_detection_trajectory}
\end{figure}

No evaluated metric detected every tested failure under the predefined rules across 233 settings (Fig.~\ref{fig:controlled_failure_detection_trajectory}).
Switching the evaluation feature extractor changed either detection status or the lowest severity detected in 105 of 279 matched CT comparisons (37.6\%) and 107 of 176 matched chest radiography comparisons (60.8\%), as shown in Supplementary Fig.~\ref{fig:supp_feature_space_detection_disagreement}. CT differences mainly concerned whether a failure was detected at all and radiographic differences the severity at which detection began. CT-FM more often detected CT intensity shifts, image-quality degradation and anatomical incoherence than MedicalNet. For radiographs, RadImageNet detected intensity shifts earlier, whereas ImageNet was more sensitive to image-quality degradation, mode dropping and anatomical incoherence. Neither representation was uniformly more sensitive.
Within a feature space, empirical-reference WAD-DIV detected duplicate collapse from the lowest tested severity in both datasets. Recall and coverage generally responded later. MedicalNet FID was less sensitive to CT duplicate collapse. Direct HU and grayscale-distribution measures were most consistently sensitive to intensity shifts. Mode dropping remained difficult to detect in both datasets.
No metric did reliably detect injected copies, while the dedicated detector detected exact-copy injection in both datasets (Supplementary Fig.~\ref{fig:supp-benchmark-detector}). Detection rarely disappeared at a stronger tested severity(Supplementary Fig.~\ref{fig:supp_feature_space_severity_curves}).
We also tested whether adding a metric to FID improved severity prediction (Supplementary Figs.~\ref{fig:supp-complementarity-full-screen} and~\ref{fig:supp-complementarity-feature-variants}). Adding WAD-DIV, an organ-intensity measure or a classifier two-sample test increased cross-validated $R^2$ by 0.641--0.836 and reduced mean absolute error by 0.128--0.192. For CT mode dropping, adding MedicalNet zero-reference WAD-DIV increased $R^2$ from -0.104 to 0.731. For CT intensity changes, adding left-lung mean-HU distributions statistic increased it from 0.124 to 0.898. \\
\textit{These findings demonstrate that model comparisons require complementary metrics with specified evaluation features: a favourable result from one measure can leave other tested failures undetected.}

\subsection{Reconstruction and generation favour different autoencoders}
\label{sec:results_selection_mismatch}

\begin{figure}[t]
\centering
\includegraphics[width=\textwidth]{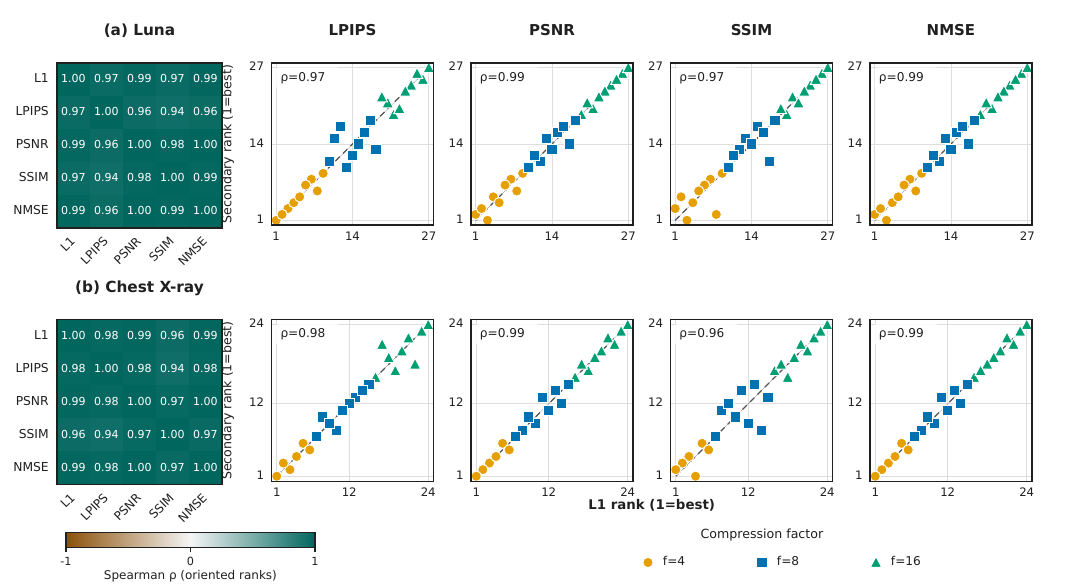}
\caption{Agreement among reconstruction metrics. Rankings of 27 CT and 24 radiographic autoencoders by mean L1, LPIPS, PSNR, SSIM and NMSE. Left panels show Spearman correlations after aligning the preferred metric directions. Remaining panels compare each metric's rank with the mean-L1 rank; rank 1 is most favourable. The dashed diagonal denotes identical rankings. Colours and symbols distinguish compression factors. All pairwise correlations exceed 0.94.}
\label{fig:supp-reconstruction-rank-map}
\end{figure}

\begin{figure}[t]
\centering
\includegraphics[width=0.8\textwidth]{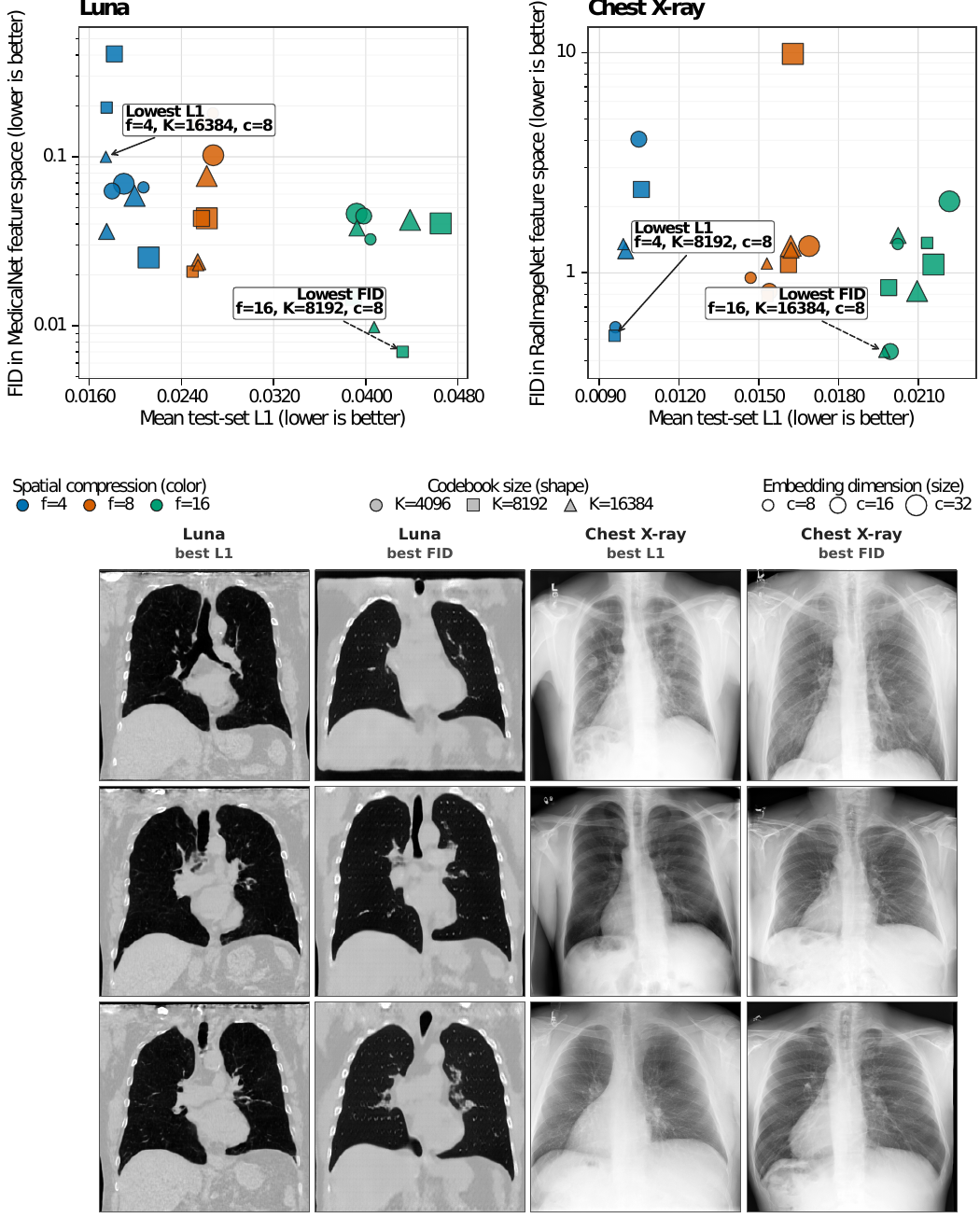}
\caption{Reconstruction and generation favour different configurations. Mean reconstruction L1 versus held-out FID for 27 CT and 24 chest X-ray pipelines. Colours indicate compression factor $f$, symbols codebook size $K$, and marker size embedding dimension $c$. Labels mark the lowest L1 and FID configurations. FID uses MedicalNet for CT and RadImageNet for chest X-ray and is shown on a logarithmic scale; values are only comparable within each dataset. Lower panels show generated examples from both selected pipelines. FID measures cohort-level quality. CT examples show central coronal slices at $[-1000,400]$ HU, selected in stored order using a fixed filter allowing at most 15\% dark perimeter pixels. Chest X-ray examples use stored indices 42--44.}
\label{fig:l1_fid_primary}
\end{figure}

The five reconstruction metrics ranked the autoencoders similarly in both datasets: all pairwise Spearman rank correlations exceeded 0.94 (Fig.~\ref{fig:supp-reconstruction-rank-map}). This agreement justified the use of mean L1, the average absolute reconstruction error, to summarise reconstruction quality.
Across all pipelines, the lowest reconstruction L1 occurred at compression factor \\ $f$ = 4, whereas the lowest primary FID occurred at $f$ = 16 (Table~\ref{tab:metric_defined_selections}; Fig.~\ref{fig:l1_fid_primary}). The reconstruction-selected CT pipeline ranked 22nd of 27 by FID. The mismatch was smaller in radiography, where the reconstruction-selected pipeline ranked third of 24. It therefore remained competitive under that metric, although it was not the rank-1 configuration.

\begin{table}[b]
\centering
\scriptsize
\setlength{\tabcolsep}{2pt}
\caption{Metric-based pipeline selection and comparisons. The top block compares the lowest-L1 and lowest-FID configurations for each dataset. Capacity and generator comparisons use the same L1-selected $f=4$ autoencoder, while HA-GAN and wavelet diffusion serve as separate complete-pipeline comparisons. L1 ranks cover all 27 CT or 24 chest X-ray configurations, while FID ranks refer to the respective autoencoder, capacity (9 models), or generator (8 models) comparison. Primary FID uses MedicalNet for CT and RadImageNet for chest X-ray. Bold values indicate the best unrounded result within each block and dataset, including ties. WAD-A measures normalised WAD-DIV agreement with the empirical reference. $N_{\mathrm{copy}}$ and $N_{\mathrm{mem}}$ count generated and training candidates above the detection threshold. Further details are provided in Supplementary Section~\ref{sec:supp-detector-methods}.}
\label{tab:metric_defined_selections}
\begin{tabular}{llrrrrrrrrrr}
\toprule
Dataset & Selection or model
& \shortstack{Mean L1 $\downarrow$\\ {[rank/$n$]}}
& \shortstack{Primary FID $\downarrow$\\ {[rank/$n$]}}
& \shortstack{FRD\\$\downarrow$} & \shortstack{Prec.\\$\uparrow$} & \shortstack{Rec.\\$\uparrow$} & \shortstack{Den.\\$\uparrow$} & \shortstack{Cov.\\$\uparrow$}
& \shortstack{WAD-A\\$\uparrow$}
& $N_{\rm copy}$
& $N_{\rm mem}$ \\
\midrule
\multicolumn{12}{@{}l}{\textit{metric-defined first-stage selection}} \\
Luna & L1-selected
& \textbf{0.0174 [1/27]} & 0.0999 [22/27] & \textbf{3.600} & \textbf{0.983} & \textbf{0.837} & 1.058 & 0.980 & 0.427 & 374 & 257 \\
Luna & FID-selected
& 0.0432 [25/27] & \textbf{0.0070 [1/27]} & 4.066 & 0.964 & 0.735 & \textbf{1.131} & \textbf{1.000} & \textbf{0.452} & 625 & 380 \\
Chest X ray & L1-selected
& \textbf{0.0096 [1/24]} & 0.5174 [3/24] & 2.693 & 0.910 & \textbf{0.751} & 0.665 & 0.780 & \textbf{0.930} & 369 & 338 \\
Chest X ray & FID-selected
& 0.0197 [16/24] & \textbf{0.4379 [1/24]} & \textbf{2.421} & \textbf{0.940} & 0.708 & \textbf{0.807} & \textbf{0.847} & 0.865 & 672 & 620 \\
\midrule
\multicolumn{12}{@{}l}{\textit{U-Net capacity at the globally L1-selected first stage}} \\
Luna & Small U-Net
& \textbf{0.0174 [1/27]} & 0.3713 [9/9] & 3.852 & 0.533 & \textbf{0.878} & 0.358 & 0.796 & \textbf{0.581} & 450 & 96 \\
Luna & Base U-Net
& \textbf{0.0174 [1/27]} & 0.0999 [8/9] & \textbf{3.600} & 0.983 & 0.837 & 1.058 & \textbf{0.980} & 0.427 & 374 & 257 \\
Luna & Wide U-Net
& \textbf{0.0174 [1/27]} & \textbf{0.0278 [6/9]} & 3.758 & \textbf{0.988} & 0.510 & \textbf{1.260} & \textbf{0.980} & 0.412 & 486 & 284 \\
\addlinespace[1pt]
Chest X ray & Small U-Net
& \textbf{0.0096 [1/24]} & 1.0932 [6/9] & 2.915 & \textbf{0.921} & 0.737 & \textbf{0.714} & 0.685 & 0.898 & 252 & 245 \\
Chest X ray & Base U-Net
& \textbf{0.0096 [1/24]} & \textbf{0.5174 [3/9]} & \textbf{2.693} & 0.910 & 0.751 & 0.665 & \textbf{0.780} & \textbf{0.930} & 369 & 338 \\
Chest X ray & Wide U-Net
& \textbf{0.0096 [1/24]} & 1.3199 [8/9] & 3.658 & 0.912 & \textbf{0.779} & 0.616 & 0.594 & 0.874 & 388 & 365 \\
\midrule
\multicolumn{12}{@{}l}{\textit{Generative architecture at the globally L1-selected first stage and end-to-end comparators}} \\
Luna & Base U-Net
& \textbf{0.0174 [1/27]} & 0.0999 [7/8] & 3.600 & 0.983 & 0.837 & 1.058 & 0.980 & 0.427 & 374 & 257 \\
Luna & DiT
& \textbf{0.0174 [1/27]} & \textbf{0.0213 [5/8]} & \textbf{3.462} & \textbf{0.991} & \textbf{0.878} & \textbf{1.267} & \textbf{1.000} & 0.469 & 518 & 403 \\
Luna & HA-GAN
& -- & 0.0425 [6/8] & 4.057 & 0.959 & 0.735 & 1.016 & \textbf{1.000} & 0.470 & 790 & 5 \\
Luna & WDM-3D
& -- & 0.2640 [8/8] & -- & 0.580 & 0.102 & 0.258 & 0.082 & \textbf{0.484} & 731 & 6 \\
\addlinespace[1pt]
Chest X ray & Base U-Net
& \textbf{0.0096 [1/24]} & 0.5174 [5/8] & 2.693 & 0.910 & 0.751 & 0.665 & 0.780 & 0.930 & 369 & 338 \\
Chest X ray & DiT
& \textbf{0.0096 [1/24]} & 0.4878 [4/8] & \textbf{2.446} & \textbf{0.933} & 0.808 & 0.787 & 0.861 & 0.918 & 373 & 386 \\
Chest X ray & HA-GAN-2D
& -- & 1.7393 [8/8] & 2.515 & 0.794 & 0.374 & 0.478 & 0.401 & 0.834 & 280 & 141 \\
Chest X ray & WaveDiff
& -- & \textbf{0.2495 [1/8]} & 3.300 & 0.927 & \textbf{0.880} & \textbf{0.828} & \textbf{0.939} & \textbf{0.975} & 326 & 341 \\
\bottomrule
\end{tabular}
\end{table}

Stronger compression increased reconstruction error but reduced average FID. Larger codebooks and more embedding channels did not consistently improve FID. Measures of how extensively the autoencoder used its codebook also did not reliably indicate which pipelines would achieve lower FID (Supplementary Figs.~\ref{fig:supp-factor-endpoints}--\ref{fig:supp-multimetric-l1}). The full baseline metrics are reported in Supplementary Tables~\ref{tab:supp-grid-Luna}--\ref{tab:supp-grid-ChestXray}. Supplementary Figs.~\ref{fig:supp-paired-reconstructions}--\ref{fig:supp-configuration-examples} show paired reconstructions and generated examples. Supplementary Fig.~\ref{fig:metric-labelled-examples} visualises generated examples alongside their cohort-level FID and radiomic-distance scores.\\
\textit{These results demonstrate that reconstruction agreement does not automatically establish downstream FID preference and the autoencoder must be assessed together with the generator.}

\subsection{Metrics and feature representations change pipeline rankings}
\label{sec:results_metric_landscape}

\begin{figure}[t]
\centering
\includegraphics[width=\textwidth]{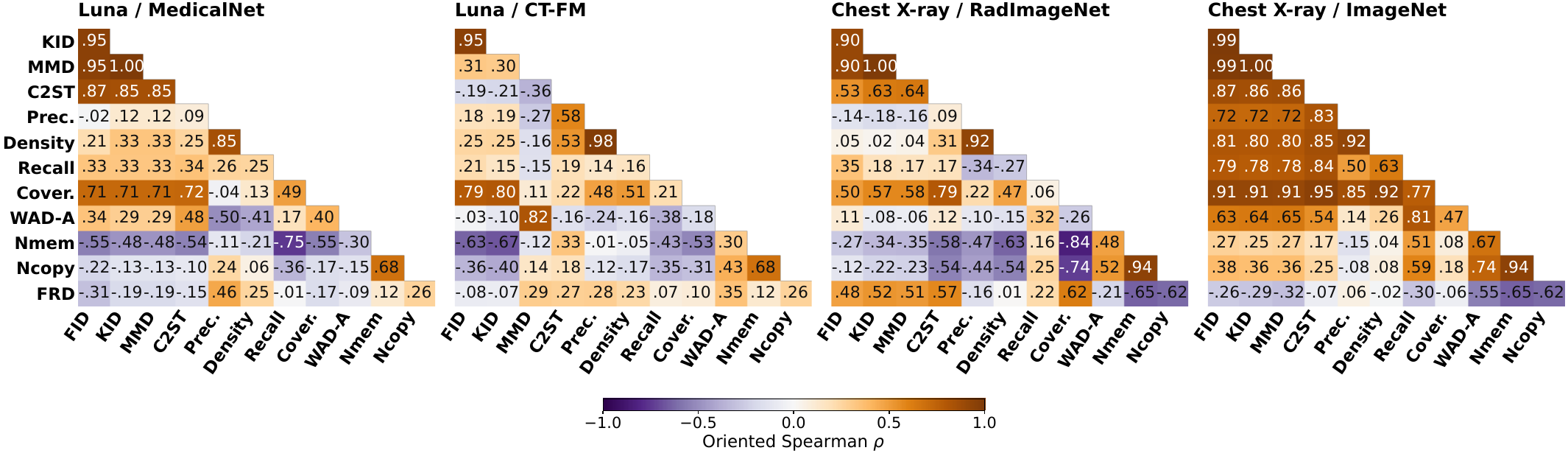}
\caption{Agreement among generative metrics depends on the evaluation setting. Spearman rank correlations across 27 CT and 24 chest X-ray pipelines, shown separately for each feature representation. Metric directions are aligned so that positive correlations indicate similar rankings. FID, KID, MMD, the C2ST gap, FRD and both detector counts are sign-reversed before calculating correlations. WAD-A measures normalised WAD-DIV agreement with the empirical reference. FRD and copy-detector counts use separate representations and are repeated across panels for comparison.}
\label{fig:generative_metric_relationship_map}
\end{figure}

The lowest-FID pipeline was not consistently preferred by other measures of the generated images (Fig.~\ref{fig:generative_metric_relationship_map}). Across the feature-based distributional metrics, rank correlations ranged from -0.502 to 1.000 for MedicalNet and from -0.379 to 0.978 for CT-FM. The corresponding X-Ray ranges were -0.344 to 0.998 for RadImageNet and 0.138 to 0.997 for Inception-v3.
The difference between distributional agreement and neighbourhood-based fidelity was particularly clear. FID and precision had almost no rank association in CT MedicalNet ($\rho$ = -0.020), a weak negative association in radiographic RadImageNet ($\rho$ = -0.140), and a stronger positive association in Inception-v3 ($\rho$ = 0.720). Their top-three sets shared no configuration under MedicalNet or RadImageNet and only one under Inception-v3. FID was more closely associated with coverage in these same representations ($\rho$ = 0.714, 0.499 and 0.915, respectively). Even under Inception-v3, FID and precision selected different configurations: (compression 8, codebook size 16384, dimension 8) and (compression 16, codebook size 16384, dimension 8).
Feature choice changed the preference without changing the metric formula. Across all radiographic pipelines, RadImageNet FID and Inception-v3 FID had rank correlation $\rho$ = 0.447. Both feature extraction settings shared one top-three configuration, but selected different rank-1 pipelines with $f$ = 16 and $f$ = 8, respectively. Among the two candidates selected by reconstruction L1 and primary FID, Inception-v3 FID favoured the reconstruction-selected $f$ = 4 pipeline. Across the full grid, however, its minimum belonged to a third configuration, (compression 8, codebook size 16384, dimension 8).
For CT, FRD was lower for the reconstruction-selected pipeline than for the MedicalNet-FID-selected pipeline (3.600 versus 4.066). For radiography, it favoured the primary-FID-selected pipeline over the L1-selected parameters (2.421 versus 2.693.  Table~\ref{tab:metric_defined_selections}).
Higher zero-reference WAD-DIV scores corresponded to greater visible variation among generated images (Supplementary Fig.~\ref{fig:wad-diversity-mosaics}). In CT, the higher-scoring model produced more varied thoracic outlines and lung shapes, while the lower-scoring model produced more similar-looking images. Radiographic examples showed differences in shape, contrast and texture. Greater variation did not necessarily coincide with better distributional agreement: the radiographic model with higher WAD-DIV also had a higher FID. WAD-DIV therefore provided complementary information about sample diversity.
We held the generated cohorts fixed and recalculated FID after replacing the held-out test references with 800 training CT volumes and 5,000 sampled training radiographs. We used MedicalNet for CT and Inception-v3 for radiography (Supplementary Fig.~\ref{fig:supp-reference-cohort}). Individual MedicalNet FID values and some model rankings changed, but overall rank agreement remained high (Spearman $\rho=0.821$). The same model ranked first, the same three models occupied the top three positions, and average FID still decreased with increasing compression. For radiography, FID values changed little, and rankings were nearly identical ($\rho=0.999$), with only two neighbouring configurations exchanging positions. Thus, the main model preferences persisted across reference cohorts, while individual CT comparisons were more sensitive to the reference used.\\
\textit{Pipeline rankings therefore depend on both the measured property and the features and reference used to measure it. Primary FID supplies one defined comparison, but not a common ordering across all evaluation criteria.}

\subsection{Medical measurements and training proximity add distinct evidence}
\label{sec:results_medical_proximity}

\begin{figure}[t]
\centering
\includegraphics[width=\textwidth]{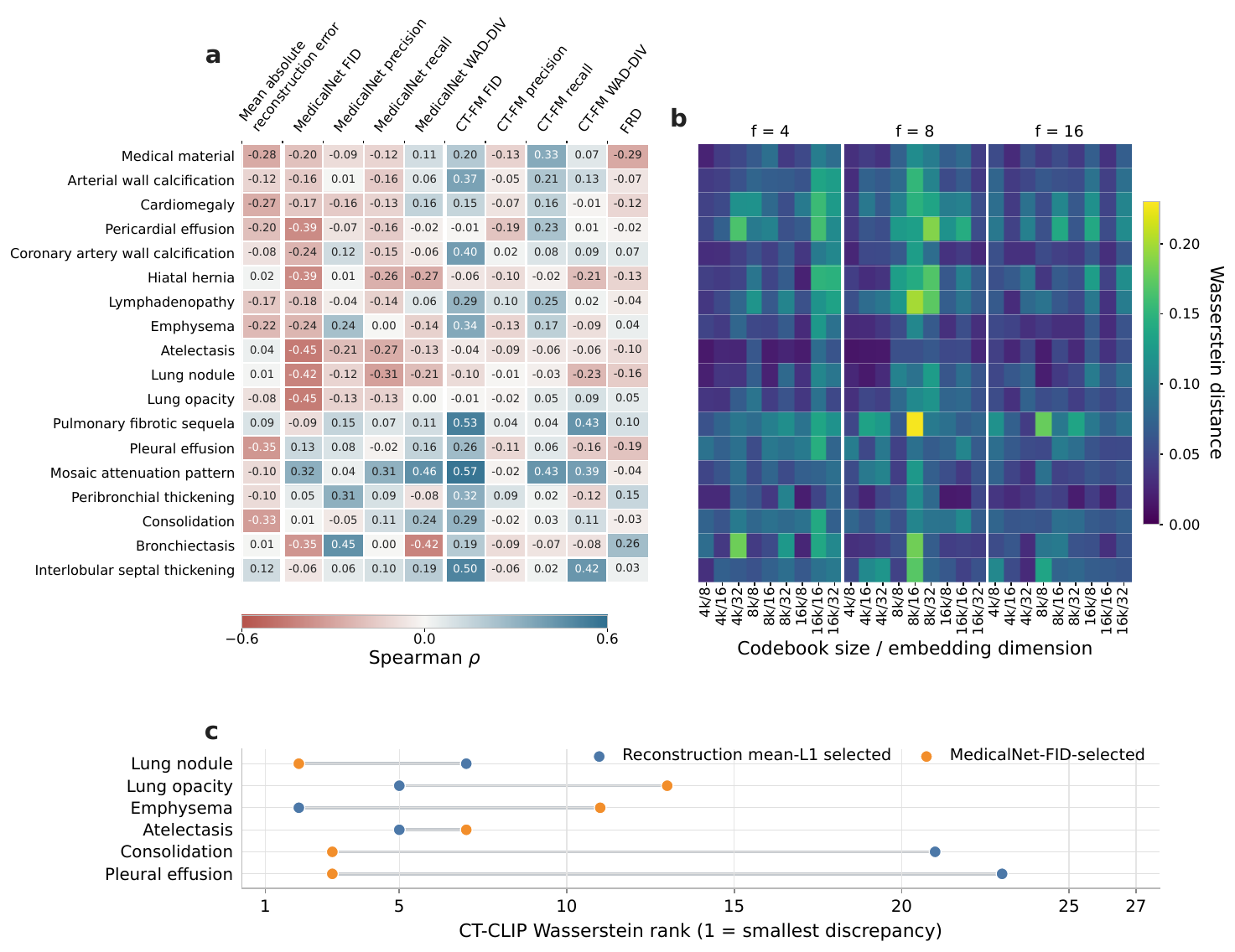}
\caption{Medical-model scores provide concept-specific comparisons. CT-CLIP score-distribution comparisons across 27 CT configurations and 18 abnormality concepts. (a) Spearman correlations of concept-specific Wasserstein distances with the listed reconstruction, feature-based and radiomic metric rankings. (b) Concept-specific distances between real and generated score distributions; lower values indicate closer agreement for that concept. (c) Ranks of the reconstruction-selected and MedicalNet-FID-selected pipelines for six example concepts; rank 1 denotes the smallest discrepancy. CT-CLIP scores \citep{hamamci2024ctclip} are model-derived proxies rather than diagnoses.}
\label{fig:ctclip_proxy_rankings}
\end{figure}

We next examined whether the feature-based rankings carried over to anatomical measurements and medical-model outputs, and whether generated images resembled individual training cases.
For the CT configurations, MedicalNet FID was weakly associated with the discrepancy between real and generated organ-volume distributions. Separate linear fits for the left lung, right lung, heart and aorta gave descriptive in-sample $R^2$ values of 0.001--0.119. Organ-volume and within-organ mean-HU distributions are shown in Supplementary Figs.~\ref{fig:supp_anatomy_volume}--\ref{fig:supp_anatomy_hu}. Failed organ segmentations were excluded when calculating the organ-volume distances.
The CT-CLIP analysis compared real and generated distributions of continuous scores for 18 abnormality concepts (Fig.~\ref{fig:ctclip_proxy_rankings}). Smaller Wasserstein distances indicated closer agreement for a given concept. Across concepts, the median rank correlation with these distances was -0.102 for reconstruction L1, -0.187 for MedicalNet FID and 0.274 for CT-FM FID. Thus, the modest median association with CT-FM did not imply consistent agreement across individual concepts.
Precision rankings transferred weakly: the median correlation was 0.009 for MedicalNet (range -0.210 to 0.449) and -0.053 for CT-FM (range -0.187 to 0.097). Median correlations for recall and WAD-DIV ranged from -0.126 to 0.058. In the direct comparison of the two primary metric-selected pipelines, the reconstruction-selected pipeline had the smaller CT-CLIP discrepancy for 12 of 18 concepts, while the MedicalNet-FID-selected pipeline did so for six.
Training-proximity counts varied across configurations (Supplementary Tables~\ref{tab:supp-grid-Luna}--\ref{tab:supp-generators}). Higher compression generally increased $N_{\mathrm{mem}}$ in both lung CT (129--271 at $f=4$ vs.\ 301--454 at $f=16$) and chest X-ray (224--364 vs.\ 499--714). FID-selected pipelines showed more detected matches than reconstruction-selected pipelines, with $N_{\mathrm{mem}}$ of 380 vs.\ 257 and $N_{\mathrm{copy}}$ of 625 vs.\ 374 for CT, and 620 vs.\ 338 and 672 vs.\ 369 for X-ray, respectively.
Generator capacity also influenced memorisation. For the reconstruction-selected first stages, $N_{\mathrm{mem}}$ increased from 96 to 284 in CT and from 245 to 365 in X-ray when moving from small to wider U-Nets. Replacing the baseline U-Net with a DiT increased counts at $f=4$ in both datasets (CT: 257 to 403; X-ray: 338 to 386), although this trend was not consistent across compression factors. Supplementary Section~\ref{sec:supp-memorisation} summarises the evaluated cohorts. Examples of detected matches are shown in Supplementary Figs.~\ref{fig:supp-closest-matches}--\ref{fig:supp-luna-closest-matches}.\\
\textit{Anatomical measurements, medical-model scores, and training proximity therefore qualify the FID comparison by revealing differences that it does not capture. However, they do not establish a single clinically superior or privacy-validated pipeline.}

\subsection{Increasing U-Net capacity does not consistently resolve the mismatch}
\label{sec:results_ldm_capacity_ablation}

We then tested whether changing the downstream U-Net could make the reconstruction-preferred autoencoder more favourable for generation. At each compression factor, we fixed the autoencoder with the lowest reconstruction L1 and compared smaller, baseline and wider U-Nets (Table \ref{tab:metric_defined_selections} and Supplementary Table~\ref{tab:supp-capacity-results}).
Changing U-Net width affected FID, but the direction depended on the dataset and compression factor. Smaller U-Nets worsened primary FID at all three compression factors in radiography and markedly worsened the CT \\ $f$ = 4 result. Wider U-Nets improved primary FID for CT at $f$ = 4 and 16, and for radiography at $f$ = 16, but worsened it in the other primary comparisons. Alternative evaluation feature extractors could also change the direction of the apparent capacity effect (Supplementary Fig.~\ref{fig:supp-capacity-fid}).
Among the three compression factors, $f$ = 16 had the lowest primary FID at every U-Net width in both datasets. The reconstruction preference for $f$ = 4 therefore remained different throughout these capacity comparisons. The lowest primary FID over the nine pipelines was achieved by the wider $f$ = 16 model in both CT (0.0057) and radiography (0.4147).

\textit{Greater U-Net capacity changes the observed metric scores, but does not automatically make the reconstruction-selected autoencoder the primary-FID preference. Improvements in FID could come with trade-offs for other metrics.}

\subsection{Changing the generator changes the FID preference}
\label{sec:results_generator_family}

We used the same reconstruction-selected autoencoder at each compression factor for both denoisers. Replacing the baseline U-Net with a diffusion transformer reduced primary FID at $f = 4$ in both datasets (Table~\ref{tab:metric_defined_selections}).
Among the three matched DiT pipelines per dataset, $f$ = 8 had the lowest primary FID. The primary-FID preference changed from $f=16$ under the baseline U-Net to $f=8$ under DiT, while reconstruction L1 favoured $f=4$ (Supplementary Table~\ref{tab:supp-generators} and Fig.~\ref{fig:study-overview}G).
HA-GAN and wavelet diffusion supplied contextual comparisons of complete pipelines. Among the eight pipelines per dataset, HA-GAN ranked sixth and WDM-3D eighth by CT MedicalNet FID. In radiography, WaveDiff had the lowest RadImageNet FID, while HA-GAN-2D ranked eighth. Examples are shown in Supplementary Fig.~\ref{fig:supp-architecture-examples}.
Other metrics favoured different pipelines. In CT, the $f$ = 4 DiT had the highest precision and density, the $f$ = 8 DiT the highest recall, and the $f$ = 8 U-Net the highest normalised empirical-reference WAD-DIV agreement. In radiography, WaveDiff led FID, recall, coverage and WAD-DIV agreement, whereas the $f$ = 8 U-Net had the highest precision and density.\\
\textit{Together, the generator comparisons support a coupled evaluation of medical-imaging LDMs. The preferred autoencoder depends on its downstream generator. While DiTs generally achieved better FID scores than U-Net-based approaches, neither generator resolved the mismatch between reconstruction quality and generative performance.}

\clearpage

\section{Discussion}
\label{sec:discussion}

Reconstruction quality alone did not identify the autoencoder preferred for unconditional medical image generation. Across the matched CT and chest-radiography LDM pipelines, reconstruction metrics largely agreed with one another, but their preferred autoencoder differed from the one with the lowest FID. A controlled benchmark explained the requirement of complementary measures: different metrics and evaluation features detected different failures. Together with the generator comparisons, these findings support assessing the autoencoder and its downstream generator as a pair, using measures chosen for the intended application.

The controlled benchmark provides a reference for interpreting generative metric responses. Its main implication is that a metric is not only a metric formula, but a complete measurement setting: the feature representation, reference cohort, preprocessing, and detection rule determine which errors become visible. No evaluated metric detected every predefined failure. A failure that is not detected in one feature space therefore cannot be assumed absent, and a favourable score cannot be interpreted as general evidence of image quality without specifying the representation in which that score was obtained.
The feature-space comparisons illustrate this dependence. Changing the evaluation feature extractor changed sensitivity. Medical-domain representations did not behave as uniformly superior versions of generic features, and secondary feature spaces did not simply confirm the primary ones. Instead, changing the representation affected the sensitivity profile: some failures became visible earlier or only in one feature space, while other failures remained difficult across representations. %
These differences should not be interpreted as solely causal effects of pretraining domain. For example, CT-FM's stronger sensitivity to selected Luna perturbations may reflect its CT-specific self-supervised pretraining and architecture \citep{projectlighter_ctfm}, which could preserve scan-level appearance or structural variation differently from a MedicalNet segmentation encoder. Similarly, ImageNet responses in chest X-ray show that a non-medical feature extractor can still be sensitive to some medically motivated perturbations.
Additionally, these findings caution against adopting generative metrics from natural-image benchmarks without validating their sensitivity to the medical failure modes of interest. Established use in computer vision alone does not demonstrate that a metric captures the differences relevant to a particular medical synthesis task.
This interpretation agrees with and extends the MRI-based stress tests of \citet{deo2025metrics}. Our results support their general principle in lung CT and chest radiography: \textbf{cohort-level feature-distribution agreement can reveal that a cohort has changed, but it does not by itself identify the mechanism of that change or establish safety, anatomical fidelity, mode coverage, or medical plausibility.}
The comparison also clarifies why metric behaviour should be interpreted within its context. \citet{deo2025metrics} found recall informative for selected class-removal experiments, whereas our recall and coverage metrics responded more clearly to duplicate collapse than to the predefined mode-dropping families. Conversely, the empirical WAD-DIV metric detected duplicate collapse from the lowest severity in both datasets, while the Vendi Score examined by \citet{deo2025metrics} showed limited sensitivity to internal duplication. This should not be read as a direct contradiction between equivalent metrics: Vendi Score and WAD-DIV measure different aspects of diversity. Rather, the comparison shows that even diversity-oriented metrics differ in their definition.

The agreement among the five reconstruction metrics is important for interpreting the central selection result. Pairwise rank correlations exceeded 0.94 in both datasets. Mean L1 was therefore a defensible predefined summary of the paired encoder--decoder task in this study. The subsequent mismatch cannot be dismissed as an arbitrary consequence of choosing one unusual reconstruction measure: \textbf{the reconstruction metrics largely agreed with one another, yet the generative quality metric favoured different regions of the first-stage design space.}
Reconstruction fidelity nevertheless remains informative: it measures errors introduced when an observed image passes through the encoder and decoder, which unpaired cohort comparisons do not assess.
A more compressed representation discards more input detail, yet may provide a lower-dimensional or more regular distribution for a finite-capacity generator to learn. Similar reconstruction--generation tensions have been reported outside the medical domain \citep{yao2025reconstruction,skorokhodov2025diffusability}, while medical LDM workflows commonly freeze a first stage selected before downstream training \citep{pinaya2022brainldm,wang2025_3d_meddiffusion,guo2025maisi, seyfarth2026medlord,khader2023meddiffusion}.
Codebook size, embedding dimension, used-code and effective-code fractions showed limited relationships with downstream generative performance. These utilisation statistics remain useful for identifying underuse or collapse. They do not, however, describe spatial arrangements, dependencies between latent locations, or the geometry of the code vectors. Their limited association with FID therefore does not imply that latent structure is unimportant. Rather, marginal utilisation statistics alone do not characterise the representation that the generator must learn.

The generated-image metrics ranked pipelines differently because they assessed different
properties, including distributional agreement, coverage and individual-image similarity. Such disagreement is informative when each metric's purpose is clear. The disagreement between FID, PRDC, and WAD-DIV should therefore not be interpreted as evidence that one metric is valid and the others are wrong. It indicates that the pipelines induced different trade-offs across distributional fidelity, support, and diversity. In fact, low correlation between metrics alone does not establish that an additional metric provides useful information. The severity-prediction analysis addresses this more directly: in selected comparisons, \textbf{FID combined with an additional metric predicted the imposed severity more accurately than FID alone} under grouped cross-validation. Conversely, close agreement between mathematically related measures, such as KID and MMD, does not provide fully independent confirmation because they summarise similar aspects of the feature distribution.
FID was the prespecified primary comparison metric for assessing whether reconstruction-based rankings transferred to generation, not a criterion for overall image quality, medical plausibility, or clinical usefulness. The chest-X-ray feature-space analysis illustrates this boundary: RadImageNet and ImageNet pretrained models could favour different configurations even though both used the same FID formula.
The medical-proxy analyses extended the same conclusion to domain-oriented readouts. In Luna, organ-volume agreement, CT-FM responses, and CT-CLIP abnormality-score distributions did not simply reproduce the MedicalNet FID ranking. CT-CLIP rankings were concept-dependent, and the reconstruction- and FID-selected pipelines were favoured for different abnormality concepts.
Such differences do not invalidate either representation. They show that a selection statement is conditional on the feature space in which it is made.
Training proximity addresses a further question: whether individual generated images closely resemble training cases. A model matching the overall feature distribution is susceptible for reproducing training cases. A cohort-level score may not reveal those matches and if patient-specific training content is reproduced, sharing synthetic images could expose sensitive information.

The generator comparisons provide direct evidence that \textbf{an autoencoder preference depends on the model trained on its representation}. Changing U-Net capacity altered the generated cohorts, but did not produce a monotonic scaling rule. Smaller models were often disadvantageous, which is consistent with the larger latent representation placing greater demands on the denoiser. Widening, however, helped only in selected compression--feature-space combinations and sometimes changed conclusions across feature representations. Thus, more capacity did not help to overcome the learnability gap.

The complementary metrics further showed why capacity cannot be interpreted solely through single distribution-based metrics. Improvements in FID could coincide with weaker support, less favourable local-distance structure, or worse intensity- and anatomy-sensitive readouts, whereas configurations with less favourable FID could retain advantages under recall, coverage, or WAD-DIV.
The generator-family analysis supported the same interpretation. Replacing the baseline U-Net with a capacity-matched DiT improved primary FID for the globally mean-L1-selected autoencoder in both datasets, showing that the downstream architecture moderated the consequence of first-stage choice. However, this improvement still did not make the reconstruction-selected first stage the FID-optimal configuration within the matched DiT comparison. The reconstruction--generation mismatch therefore persisted beyond the baseline U-Net setting, although its magnitude depended on the generator family.
The HA-GAN and wavelet-diffusion models provide a broader contextual view. They reinforce the metric-dependent nature of model assessment.

This study comprises 124 trained models, 233 controlled benchmark settings, and 918 checkpoint evaluations. The model comparisons are descriptive and use one training realisation and one fixed generated cohort per condition. Its purpose was to examine the limits of reconstruction-based first-stage selection and single-metric evaluation. Exhaustive comparisons across autoencoder and generator families and the development of new metrics were outside this scope.
The controlled benchmark covers a broad but finite set of independently imposed failures and cannot reproduce all errors or interactions that may arise in trained generators.
These findings motivate future methodological work on generative medical-image evaluation. While complementary metrics provide information beyond reconstruction fidelity and FID, they also increase evaluation effort and complexity. Future work should therefore clarify what metrics measure, assess their relevance to downstream tasks and clinical judgements, and explore whether new theoretical concepts or more informative metrics can reduce this burden. The longer-term objective is a unified, reproducible evaluation framework that links intended use to complementary measurements and transparent reporting, rather than imposing a single overall quality score.

For medical image generation, we recommend defining the intended use and relevant failure modes before choosing the metrics, feature representations and reference cohorts. Controlled benchmarks can inform these choices. Combining results into a single score requires a justified rule for weighting the different properties. For latent diffusion, reconstruction quality can guide candidate screening. We then recommend evaluating each shortlisted autoencoder with the intended downstream generator under a common protocol. Reporting reconstruction, distibutional agreement and medically relevant measurements seperately allows readers to see which properties improve and where the choice of pipeline involves a trade-off.

\section*{Competing interests}

S.E. has received honoraria from Abbott and Boehringer Ingelheim. The other authors declare no competing interests.

\section*{Funding}

This work was supported by the Heidelberg Faculty of Medicine at Heidelberg University, the Carl Zeiss Foundation through the Multi-DimensionAI project (P2022-08-010), the EU Horizon project DVPS (grant 101213369), the Klaus Tschira Foundation, and the DZHK (reference number 81X3500151, project number 2098).

\section*{Data availability}

The 888 CT scans analysed in this study are available from the LUNA16 challenge~\citep{SETIO20171} through two Zenodo deposits (\href{https://doi.org/10.5281/zenodo.2595812}{10.5281/zenodo.2595812} and \href{https://doi.org/10.5281/zenodo.2596478}{10.5281/zenodo.2596478}). The fixed subset of 20,000 radiographs was drawn from ChestX-ray14, which was provided by the NIH Clinical Center and is available from the \href{https://nihcc.app.box.com/v/ChestXray-NIHCC}{original NIH download site}~\citep{wang2017chestxray8}. Supplementary Table~\ref{tab:supp-training} gives the patient-disjoint split sizes. The metrics for the two primary model grids are reported in Supplementary Tables~\ref{tab:supp-grid-Luna} and~\ref{tab:supp-grid-ChestXray}.

\section*{Code availability}

Code for model training, evaluation and the controlled metric benchmark is publicly available at \url{https://github.com/Cardio-AI/medical-latent-diffusion-evaluation}.

\section*{Ethics statement}

This study used publicly released, de-identified CT scans from LUNA16 and chest radiographs from NIH ChestX-ray14 for secondary analysis. The study involved no participant recruitment or access to identifiable clinical records.

\section*{Acknowledgements}

We thank SDS@hd for providing data storage. The service is supported by the Ministry of Science, Research and the Arts Baden-W\"urttemberg (MWK) and the German Research Foundation (DFG) through grants INST~35/1314-1~FUGG and INST~35/1503-1~FUGG. We acknowledge computing resources provided by the State of Baden-W\"urttemberg through bwHPC and supported by the DFG through grant INST~35/1597-1~FUGG.

\section*{Author contributions}

M.S.\ conceived and designed the study, conducted the experiments, performed the evaluations and wrote the initial manuscript draft. Y.F.\ helped interpret the results and contributed to writing and revising the manuscript. S.E.\ and S.U.H.D.\ helped develop and refine the study concept, contributed to interpreting the results and critically reviewed the manuscript.

\clearpage
\input{supplement.tex}
\end{document}

%% file: supplement.tex
\newgeometry{margin=18mm}
\begingroup
\small
\setcounter{section}{0}
\setcounter{figure}{0}
\setcounter{table}{0}
\renewcommand*{\theHsection}{supp.\arabic{section}}
\renewcommand*{\theHfigure}{supp.\arabic{figure}}
\renewcommand*{\theHtable}{supp.\arabic{table}}
\definecolor{suppblue}{RGB}{38,68,89}
\definecolor{panelgray}{RGB}{246,248,250}
\captionsetup{font=small,labelfont=bf,skip=5pt}
\setlength{\parindent}{0pt}
\setlength{\parskip}{4pt}
\setlength{\emergencystretch}{2em}
\setlength{\tabcolsep}{4pt}
\renewcommand{\arraystretch}{1.08}
\renewcommand{\thesection}{S\arabic{section}}
\renewcommand{\thefigure}{S\arabic{figure}}
\renewcommand{\thetable}{S\arabic{table}}
\setcounter{tocdepth}{1}
\pagestyle{fancy}
\fancyhf{}
\fancyhead[L]{\small Supplementary Information}
\fancyhead[R]{\small Metric sensitivity and autoencoder selection}
\fancyfoot[C]{\thepage}
\setlength{\headheight}{14pt}
\renewcommand{\headrulewidth}{0.3pt}

\newcommand{\reservedpanel}[3]{%
  \begin{figure}[H]\centering
  \fcolorbox{suppblue}{panelgray}{\begin{minipage}[c][#1][c]{0.92\linewidth}
  \centering\color{suppblue}\large\textbf{Image panel reserved}\\[12pt]
  \normalsize #2
  \end{minipage}}
  \caption{#3}\end{figure}}

\thispagestyle{plain}
{\noindent\large\bfseries Supplementary Information\par}
\vspace{8pt}
{\raggedright\LARGE\bfseries The limits of reconstruction fidelity and evaluation in medical image synthesis with latent diffusion\par}
\vspace{8pt}
Marvin Seyfarth, Yannik Frisch, Sandy Engelhardt, Salman Ul Hassan Dar
\vspace{10pt}

\FloatBarrier
\section{Supplementary methods}
\subsection{Data and primary training protocol}

\begin{table}[H]\centering\small
\caption{Dataset splits and primary training settings.}\label{tab:supp-training}
\begin{tabularx}{\linewidth}{lXX}\toprule
 & LUNA16 CT & NIH ChestX-ray14 subset \citep{wang2017chestxray8} \\\midrule
Training / validation / test & 800 / 39 / 49 & 10,000 / 5,000 / 5,000 \\
VQ-GAN batch size & 2 & 5 \\
Patch size by epoch & $64^3$ (0--49), $128^3$ (50--99), $192\times192\times128$ (100--199) & $256^2$ (0--99), $512^2$ (100--199), $1024^2$ (200--299) \\
Perceptual / adversarial weights & 0.005 & 0.01 \\
Linear loss-weight warm-up & 10 epochs & 5 epochs \\
Primary U-Net batch size & 1 & 20 \\
Primary U-Net updates & 300,000 & 200,000 \\\bottomrule
\end{tabularx}\end{table}

The VQ-GAN generator and discriminator used separate AdamW optimisers with learning rates of $5\times10^{-5}$ and $2\times10^{-5}$, respectively. Both rates decayed by a factor of 0.999 per epoch.

The primary U-Net used two residual blocks per level and attention at the two deepest levels, with 32 channels per head. The input and output each had $c$ channels. AdamW used a learning rate of $2.5\times10^{-5}$ with the same decay. Generation used 300 DDPM steps \citep{seyfarth2026medlord}.

\subsection*{Checkpoint evaluation}
The autoencoder curves in Fig.~\ref{fig:supp-training-behaviour} use the validation sets in Table~\ref{tab:supp-training}. We generated samples for FID and KID with 50 DDIM steps, $\eta=0$ and matched initial noise. CT FID and KID use MedicalNet with 200 generated and 39 validation volumes. The radiographic scores use Inception-v3 with 500 generated and 500 validation images.
\begin{figure}[H]
\centering
\includegraphics[width=\linewidth,height=0.73\textheight,keepaspectratio]{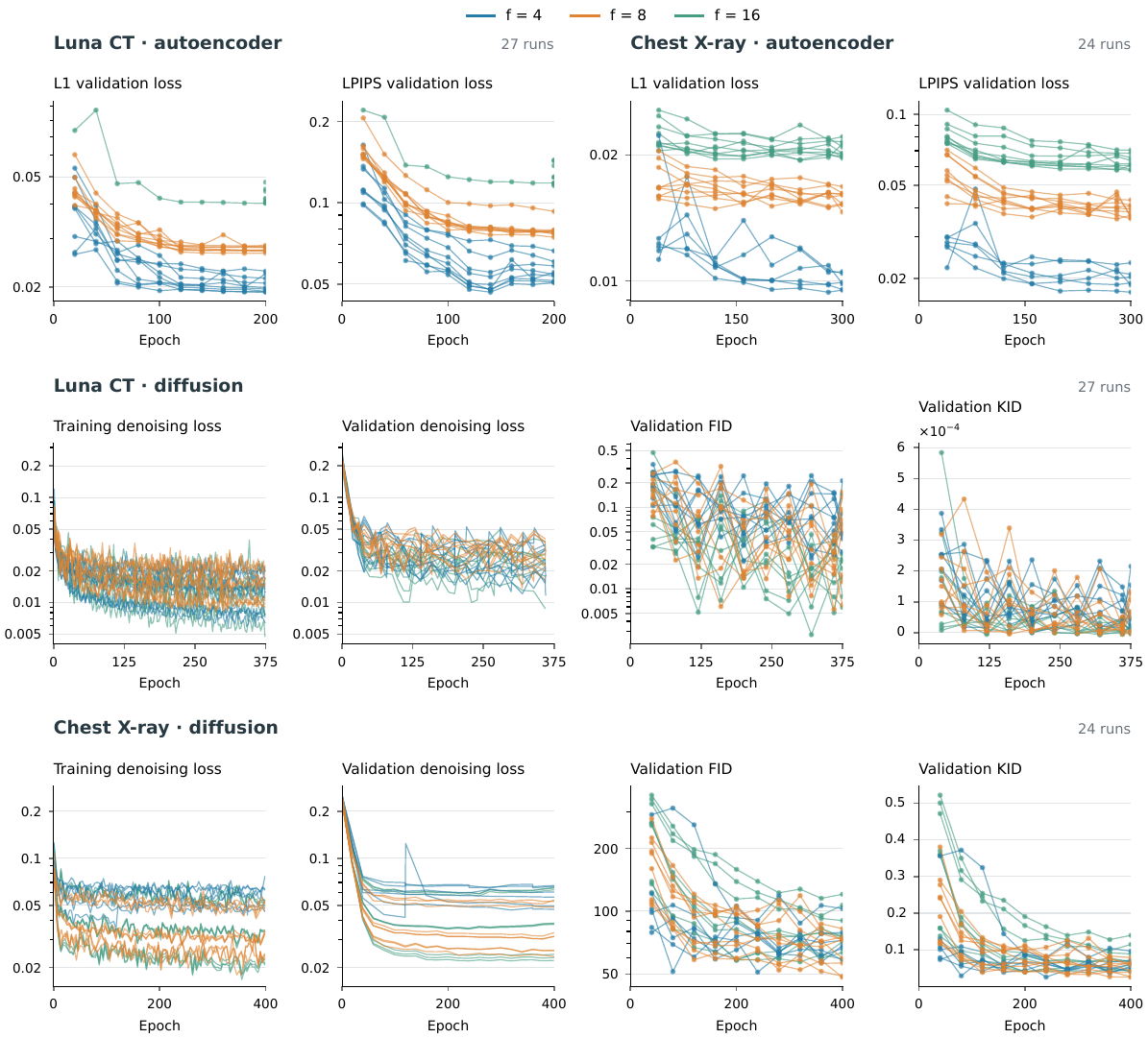}
\caption{Training curves. The top row shows autoencoder validation L1 and LPIPS. The lower rows show diffusion losses, FID and KID. Lines show individual runs and colours show compression factors. Dots show the final measurements for eight CT autoencoders. Training losses are medians in 2,000-update bins. Validation points show individual measurements. Axes are logarithmic except for KID.}
\label{fig:supp-training-behaviour}
\end{figure}

\subsection{Controlled metric benchmark}\label{sec:supp-benchmark-methods}

The reference, source and donor pools contained 267, 267 and 266 CT volumes and 3,334, 3,333 and 3,333 radiographs, respectively. The frozen manifest contains 153 CT and 80 radiographic settings, including subtype-specific identity conditions.

MedicalNet and CT-FM were compared on the same CT medical-proxy cohorts. Core seeds were 41001--41020 and proxy seeds were 51001--51010. Transformation seeds were fixed for each source case across replicates. Clean real--real controls used 20 replicates of 100 CT volumes or 1,024 radiographs.

In Tables~\ref{tab:supp-benchmark-transformations} and~\ref{tab:supp-benchmark-transformations-chest}, zero or identity leaves the source unchanged. Intensity gains and fractions are unitless. Signed changes were analysed separately by direction.

\begin{table}[H]
\centering
\footnotesize
\setlength{\tabcolsep}{3.0pt}
\caption{CT perturbation settings.}
\label{tab:supp-benchmark-transformations}
\begin{tabular}{>{\raggedright\arraybackslash}p{0.15\textwidth}>{\raggedright\arraybackslash}p{0.18\textwidth}>{\raggedright\arraybackslash}p{0.54\textwidth}}
\toprule
Failure family & Transformation & Identity and severity levels \\
\midrule
Global intensity & HU offset & $0,\ \pm25,\ \pm50,\ \pm100,\ \pm200$ HU \\
Global intensity & Contrast about 0 HU & $1.0,\ 0.9,\ 0.75,\ 1.1,\ 1.25$ \\
Global intensity & HU-window narrowing &
$[-1000,1000]$, $[-900,900]$, $[-750,750]$,
$[-500,500]$, and $[-250,250]$ HU \\
Local degradation & Gaussian blur & $0,\ 0.5,\ 1,\ 2,\ 4$ mm \\
Local degradation & Gaussian noise SD & $0,\ 0.01,\ 0.025,\ 0.05,\ 0.1$ of normalised range \\
Local degradation & High-frequency gain & $1,\ 0.75,\ 0.5,\ 0.25,\ 0$; 1-mm low-pass scale \\
Local degradation & Central checkerboard &
Amplitude $0,\ 0.025,\ 0.05,\ 0.1,\ 0.2$; central 25\% per axis; blocks $4{\times}16{\times}16$ voxels \\
Support loss & Remove auxiliary strata & Removed fraction $0,\ 0.25,\ 0.5,\ 0.75,\ 1$ from largest or smallest eligible cluster \\
Diversity collapse & Reuse unique cases & Unique fraction $1,\ 0.9,\ 0.75,\ 0.5,\ 0.25$ \\
Donor-copy injection & Exact or near copy &
Injected fraction $0,\ 0.01,\ 0.05,\ 0.1,\ 0.25,\ 0.5$; near copy uses a one-voxel shift per axis, $+5$ HU, and 2-HU noise SD \\
Volumetric anatomy & Slice repetition & Affected fraction $0,\ 0.05,\ 0.1,\ 0.2,\ 0.4$ \\
Volumetric anatomy & Organ scaling &
Absolute scale change $0,\ 0.05,\ 0.1,\ 0.2,\ 0.3$, contraction and expansion, for left/right lung, heart, and aorta \\
Volumetric anatomy & Organ HU offset & $0,\ \pm25,\ \pm50,\ \pm100,\ \pm200$ HU for each retained organ \\
\bottomrule
\end{tabular}
\end{table}

For CT mode dropping, we built auxiliary strata from z-scored $16^3$ HU grids using 16-component PCA and k-means with six clusters. The radiographic strata used NIH metadata. The resulting mode-dropping cohorts contained unique cases.

\begin{table}[H]
\centering
\footnotesize
\setlength{\tabcolsep}{3.0pt}
\caption{Chest X-ray perturbation settings.}
\label{tab:supp-benchmark-transformations-chest}
\begin{tabular}{>{\raggedright\arraybackslash}p{0.15\textwidth}>{\raggedright\arraybackslash}p{0.18\textwidth}>{\raggedright\arraybackslash}p{0.54\textwidth}}
\toprule
Failure family & Transformation & Identity and severity levels \\
\midrule
Global intensity & Brightness offset & $0,\ \pm0.025,\ \pm0.05,\ \pm0.1,\ \pm0.2$ \\
Global intensity & Gamma & $1.0,\ 0.9,\ 0.75,\ 0.5,\ 1.1,\ 1.25,\ 1.5$ \\
Global intensity & Contrast about 0.5 &
$1.0$, $0.9$, $0.75$, $0.5$, $0.25$, $1.1$, $1.25$, $1.5$, and $1.75$ \\
Local degradation & Gaussian blur & $\sigma = 0,\ 0.5,\ 1,\ 2,\ 4$ mm \\
Local degradation & Gaussian noise SD & $0,\ 0.01,\ 0.025,\ 0.05,\ 0.1$ of normalised range \\
Local degradation & High-frequency gain & $1,\ 0.75,\ 0.5,\ 0.25,\ 0$; Gaussian low-pass $\sigma = 2$ mm \\
Local degradation & Central checkerboard &
Amplitude $0,\ 0.025,\ 0.05,\ 0.1,\ 0.2$; central 25\% per axis; $16{\times}16$-pixel blocks \\
Support loss & Remove metadata strata & Removed fraction $0,\ 0.25,\ 0.5,\ 0.75,\ 1$ from common no-finding or rare cardiomegaly strata \\
Diversity collapse & Reuse unique cases & Unique fraction $1,\ 0.9,\ 0.75,\ 0.5,\ 0.25$ \\
Donor-copy injection & Exact or near copy &
Injected fraction $0,\ 0.01,\ 0.05,\ 0.1,\ 0.25,\ 0.5$; near copy uses a one-pixel shift per axis, $+0.005$ intensity, and 0.002 noise SD \\
Thoracic anatomy & Elastic deformation & Displacement $0,\ 4,\ 8,\ 16,\ 32$ pixels in a fixed central thoracic ellipse \\
\bottomrule
\end{tabular}
\end{table}

We applied both Gaussian operations to the min--max-normalised PNGs before feature-extractor resizing. To convert the nominal standard deviations from millimetres to pixels, we assumed the $1024\times1024$ PNGs covered the original field of view. For each axis, we multiplied the original pixel spacing from the NIH metadata by the original-to-PNG dimension ratio. We divided the nominal scale by this effective spacing.

\textbf{Severity coding.} The stored ranks are ordinal. CT contrast uses ranks 0, 2 and 4 in each direction. Radiographic gamma uses ranks 0, 1, 3 and 4. Donor fractions of 0\%, 1\%, 5\%, 10\%, 25\% and 50\% map to ranks 0, 1, 2, 2, 3 and 4, respectively.

\textbf{Severity prediction.} We ran the screen separately for each evaluation cohort. We calculated $R^2$ and mean absolute error (MAE) from pooled held-out predictions.

\subsection{Feature-extractor preprocessing}

\begin{table}[H]\centering\small
\caption{Feature-extractor inputs and outputs.}\label{tab:supp-extractor-preprocessing}
\begin{tabularx}{\linewidth}{lX}\toprule
Representation & Input preparation and output \\\midrule
MedicalNet & After primary CT preprocessing, z-score each volume and resize to $112^3$ by trilinear interpolation. Extract 3-D ResNet-50 average-pool features. \\
CT-FM & Clip to $[-1024,2048]$ HU and scale to $[0,1]$. Resample to 1.5-mm isotropic spacing and crop or pad to $128^3$. Extract 512 features. \\
CT-CLIP & Clip to $[-1000,1000]$ HU and resample to $1.5\times0.75\times0.75$ mm in $z,y,x$. Prepare volumes at $480\times480\times240$ (height, width, depth) and extract concept scores. \\
RadImageNet & Scale grayscale to $[0,1]$ and repeat across RGB channels. Resize to $224^2$ by antialiased bilinear interpolation. Apply ImageNet normalisation and extract ResNet-50 pooled features. \\
Inception-v3 & Scale grayscale to $[0,1]$ and repeat across RGB channels. Use the released preprocessing and pool3 features. \\\bottomrule
\end{tabularx}\end{table}
For MedicalNet, HA-GAN and WDM-3D CT outputs retain their native $256^3$ field of view.

\subsection{Architecture comparisons and analysis definitions}\label{sec:supp-architecture}
Capacity and DiT comparisons use the following fixed configurations $(f,K,c)$:
\begin{itemize}
\setlength{\itemsep}{0pt}
\item CT: $(4,16384,8)$, $(8,8192,8)$ and $(16,8192,16)$.
\item Chest X-ray: $(4,8192,8)$, $(8,4096,8)$ and $(16,16384,8)$.
\end{itemize}
The wider CT $f=4$ U-Net used channels [64, 160, 320, 640] and two-rank fully sharded data parallelism to fit memory. The smaller and wider CT models used 375 epochs, an effective batch size of 2 and 150,000 updates. The baseline settings are in Table~\ref{tab:supp-training}. All radiographic widths used the same training schedule as the baseline.

\textbf{Grid comparisons.} Top-three sets retain ties. We tracked the largest absolute rank gaps between FID and PRDC. Factor plots use unweighted means. Balanced radiographic comparisons of $f$ and $K$ use $c\in\{8,16\}$. Comparisons of $c$ use $f\in\{8,16\}$. The used-code fraction is $(K-D)/K$, where $D$ counts unused entries.

\textbf{Medical proxies.} TotalSegmentator supplied the masks \citep{wasserthal2023totalsegmentator}. Each trained-grid comparison used 49 real and 100 generated CT volumes. Generated volumes were assigned the approximate training spacing ($0.7\times0.7\times1.25$ mm) in memory. The controlled benchmark records segmentation failures and uses valid masks for organ-volume and mean-HU summaries.

\textbf{Direct intensity measures.} Each case contributed equally to the mean probability histogram. We used 2,001 bin edges from $-1000$ to $1000$ HU for CT and 257 bin edges from 0 to 1 for radiography. The Kolmogorov--Smirnov statistic is the maximum difference between the cumulative histograms. The Wasserstein distance is the integral of their absolute difference.

\subsection{Training-proximity detector}\label{sec:supp-detector-methods}
The detector thresholds were $0.8932863485426732$ for CT (8 September 2026 release) and approximately 0.4607 for radiography (7 September 2026 release) \citep{dar2026memorization}. The detector compared 800 generated with 800 training CT volumes and 5,000 generated with 10,000 training radiographs.

$N_{\mathrm{copy}}$ counted generated images whose maximum Pearson correlation with any training image exceeded a fixed threshold, thereby characterising the number of synthetic samples that were patient data copies. $N_{\mathrm{mem}}$ counted training images whose maximum correlation with any generated image exceeded the same threshold, which are the patient data samples classified as memorised. For each dataset, the threshold was calibrated at the 95th percentile of training-to-nearest-validation correlations and then held fixed across configurations.

\clearpage
\subsection{Evaluation metrics}

\begingroup
\small
\setlength{\tabcolsep}{4pt}
\renewcommand{\arraystretch}{1.12}

\begin{longtable}{@{}>{\raggedright\arraybackslash}p{0.27\linewidth}>{\raggedright\arraybackslash}p{0.69\linewidth}@{}}

\caption{Evaluation metrics and preferred directions.}
\label{tab:metric-overview}\\

\toprule
Metric / direction & What it measures \\
\midrule
\endfirsthead

\multicolumn{2}{@{}l}{\tablename~\thetable\ (continued)}\\
\toprule
Metric / direction & What it measures \\
\midrule
\endhead

\midrule
\multicolumn{2}{r@{}}{\footnotesize Continued on next page}\\
\endfoot

\bottomrule
\endlastfoot

L1 \ensuremath{\downarrow} & Mean absolute pixel or voxel error. \\

LPIPS \ensuremath{\downarrow} & Reconstruction error in learned features \citep{zhang2018lpips}. \\

PSNR \ensuremath{\uparrow} & Logarithmic score derived from mean squared error. \\

SSIM \ensuremath{\uparrow} & Similarity of local luminance, contrast and structure \citep{wang2004ssim}. \\

NMSE \ensuremath{\downarrow} & Squared reconstruction error divided by the reference image's mean squared intensity. \\

\midrule

FID \ensuremath{\downarrow} & Fr\'echet distance from feature means and covariances \citep{heusel2017ttur}. \\

KID \ensuremath{\downarrow} & Squared maximum mean discrepancy with a polynomial kernel \citep{binkowski2018mmd}. \\

MMD \ensuremath{\downarrow} & Discrepancy between kernel mean embeddings \citep{gretton2012mmd}. \\

C2ST gap \ensuremath{\downarrow} & Classifier separability, $\lvert\text{balanced accuracy}-0.5\rvert$ \citep{lopezpaz2017c2st}. \\

FRD \ensuremath{\downarrow} & Fr\'echet distance between radiomic feature distributions, reported as $\log(d_{\mathrm{F}})$ \citep{konz2026frd}. \\

\midrule

Precision \ensuremath{\uparrow} & Fraction of generated samples inside estimated real feature-space support. \\

Recall \ensuremath{\uparrow} & Fraction of real samples inside estimated generated feature-space support. \\

Density \ensuremath{\uparrow} & Normalised count of real-data neighbourhoods containing generated samples. \\

Coverage \ensuremath{\uparrow} & Fraction of real-data neighbourhoods containing a generated sample. PRDC definitions follow \citet{naeem2020prdc}. \\

\midrule

Empirical WAD-DIV \ensuremath{\downarrow} & Wasserstein distance between real and generated nearest-neighbour distance distributions \citep{seyfarth2026waddiv}. \\

WAD-A \ensuremath{\uparrow} & Normalised agreement with the empirical reference \citep{seyfarth2026waddiv}. \\

Zero-reference WAD-DIV & Feature-space spread relative to zero distance, compared with the clean real-cohort value \citep{seyfarth2026waddiv}. \\

\midrule

N\textsubscript{copy} / N\textsubscript{mem} & Counts of generated images classified as patient data copies and training images classified as memorised, respectively, at the fixed detector threshold (Section~\ref{sec:supp-detector-methods}; \citealp{dar2026memorization}). \\

\midrule

Intensity Wasserstein \ensuremath{\downarrow} & Distance between HU or grayscale distributions. \\

Organ-volume / mean-HU Wasserstein \ensuremath{\downarrow} & Distances between organ-volume or mean-HU distributions. \\

KS \ensuremath{\downarrow} & Largest difference between cumulative distributions, including left-lung mean HU. \\

CT-CLIP Wasserstein \ensuremath{\downarrow} & Distance between continuous abnormality-score distributions for each concept \citep{hamamci2024ctclip}. \\

\end{longtable}

\noindent\footnotesize
Finite-sample KID and MMD estimates can be slightly negative.

\endgroup

\textbf{Kernel metrics and C2ST.}
KID and MMD used unbiased estimates of squared MMD for the updated RadImageNet and CT-FM model grids and the MedicalNet, RadImageNet and Inception-v3 benchmarks. For $d$-dimensional feature vectors $x,y$, KID used $(1+\langle x,y\rangle/d)^3$ and MMD used $\exp(-\lVert x-y\rVert^2/d)$. C2ST used seed 0 and pooled held-out predictions to calculate balanced accuracy, with half credit for ties.

\textbf{PRDC neighbourhoods.}
We used $k=8$ for the updated RadImageNet grid, $k=6$ for the CT-FM model grid and $k=4$ for the CT-FM benchmark. For other benchmark feature sets, calibration used alternating halves of the real reference cohort. We chose the smallest tested $k$ that gave real--real precision and recall of at least 0.95. The largest tested value served as the fallback. The primary MedicalNet and secondary Inception-v3 model grids use archived scores with unverified $k$ and implementation versions.

\begin{samepage}
\textbf{Fr\'echet radiomic distance.}
Whole-image FRDv1 used the standard 2-D and 3-D configurations \citep{konz2026frd}. These extract first-order statistics and selected grey-level co-occurrence, run-length, size-zone and neighbouring grey-tone difference features. Features use original, Laplacian-of-Gaussian ($\sigma=2,3,4,5$) and wavelet images. Settings were intensity normalisation with scale 100, bin width 5, voxel array shift 300 and B-spline resampling. CT volumes were resampled to 2-mm isotropic spacing after restoring the generated volumes' working spacing (Section~\ref{sec:supp-architecture}). Radiographs used forced 2-D extraction at two-pixel spacing on the archived grid. We standardised both cohorts using held-out real statistics and kept all images. We retained feature columns with finite values across both cohorts. We computed Gaussian means and covariances and reported $\log(d_{\mathrm{F}})=\tfrac{1}{2}\log(d_{\mathrm{F}}^2)$. Versions were \texttt{frd-score} 1.0.1, PyRadiomics 3.1.1.dev111+g8ed579383 and SimpleITK 2.5.6.\par
\end{samepage}

\FloatBarrier
\section{Controlled benchmark results}
\begin{figure}[H]\centering
\includegraphics[width=\linewidth,height=0.33\textheight,keepaspectratio]{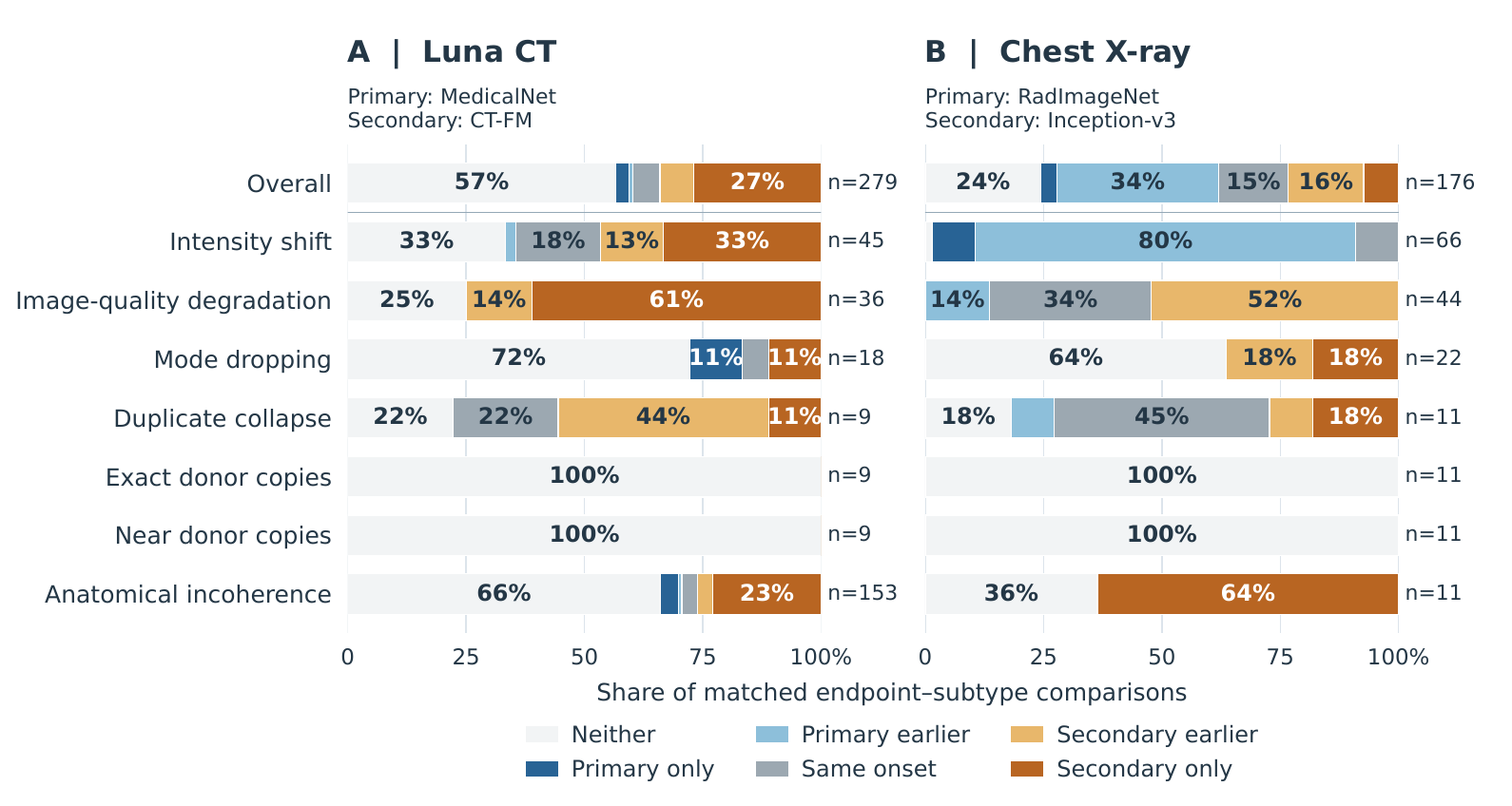}
\caption{Detection agreement between feature extractors. Detection status or onset changed in 105 of 279 matched CT comparisons and 107 of 176 radiographic comparisons.}\label{fig:supp_feature_space_detection_disagreement}\end{figure}

\begin{figure}[H]\centering
\includegraphics[width=\linewidth,height=0.36\textheight,keepaspectratio]{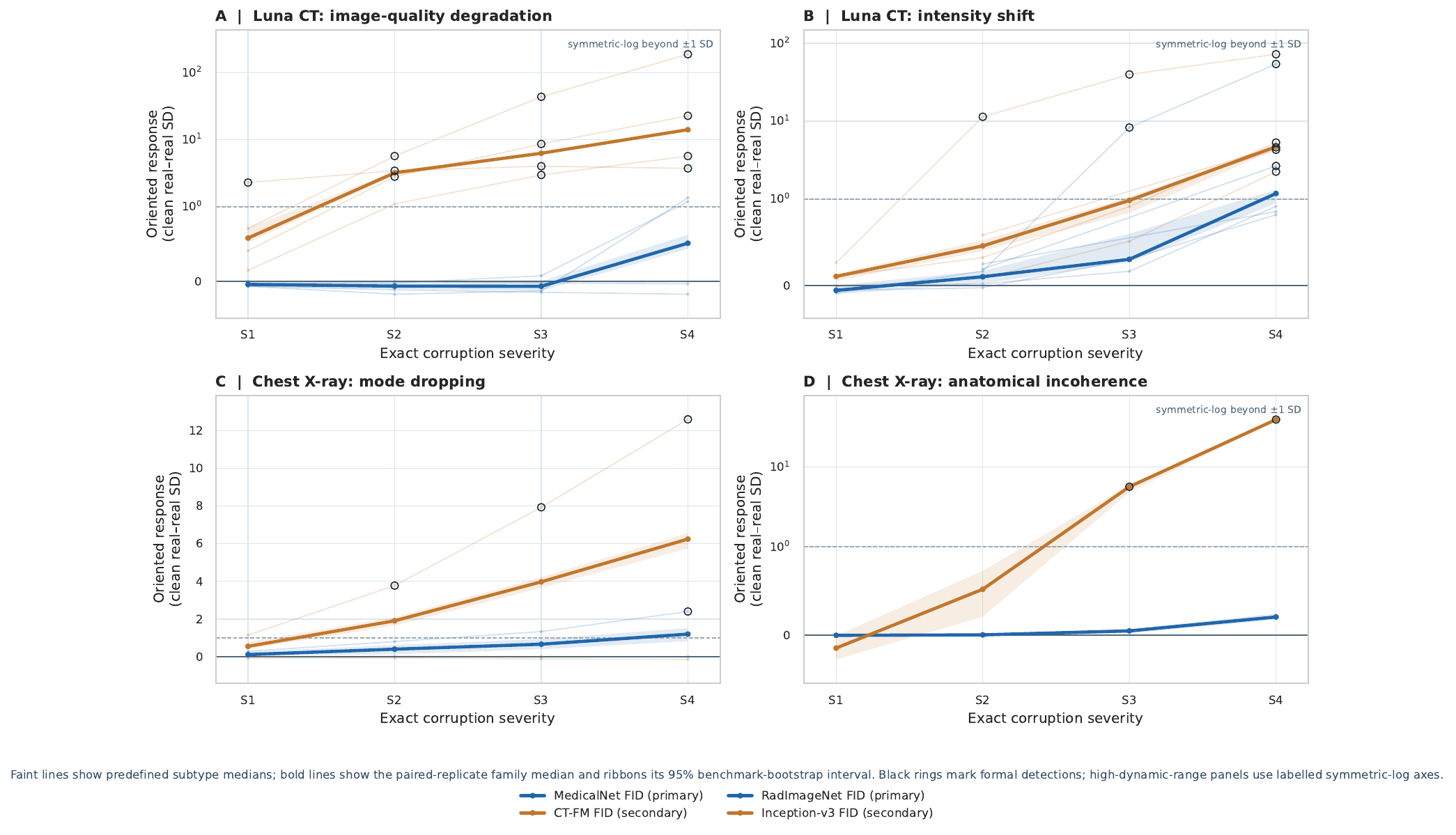}
\caption{Severity-response profiles. Paired changes are scaled by the clean real--real standard deviation. Faint lines show subtype medians. Bold lines and bands show family medians and 95\% bootstrap intervals. Rings mark detections.}\label{fig:supp_feature_space_severity_curves}\end{figure}

\begin{figure}[H]\centering
\includegraphics[width=\linewidth,height=0.36\textheight,keepaspectratio]{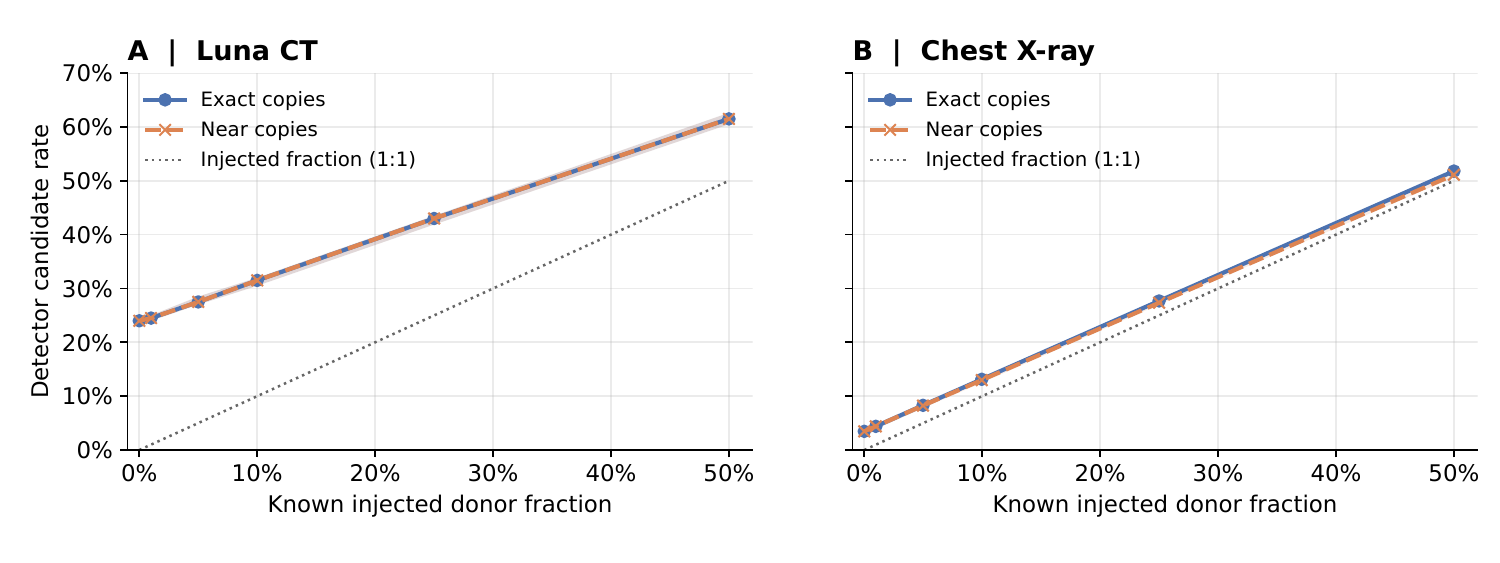}
\caption{Detector response to donor-copy injection. Lines show median rates of cases classified as donor copies across 20 replicates. The detector searches the donor pool. The diagonal shows the injected fraction. Unchanged cases can also cross the threshold.}\label{fig:supp-benchmark-detector}\end{figure}

\begin{figure}[H]\centering
\includegraphics[width=\linewidth,height=0.46\textheight,keepaspectratio]{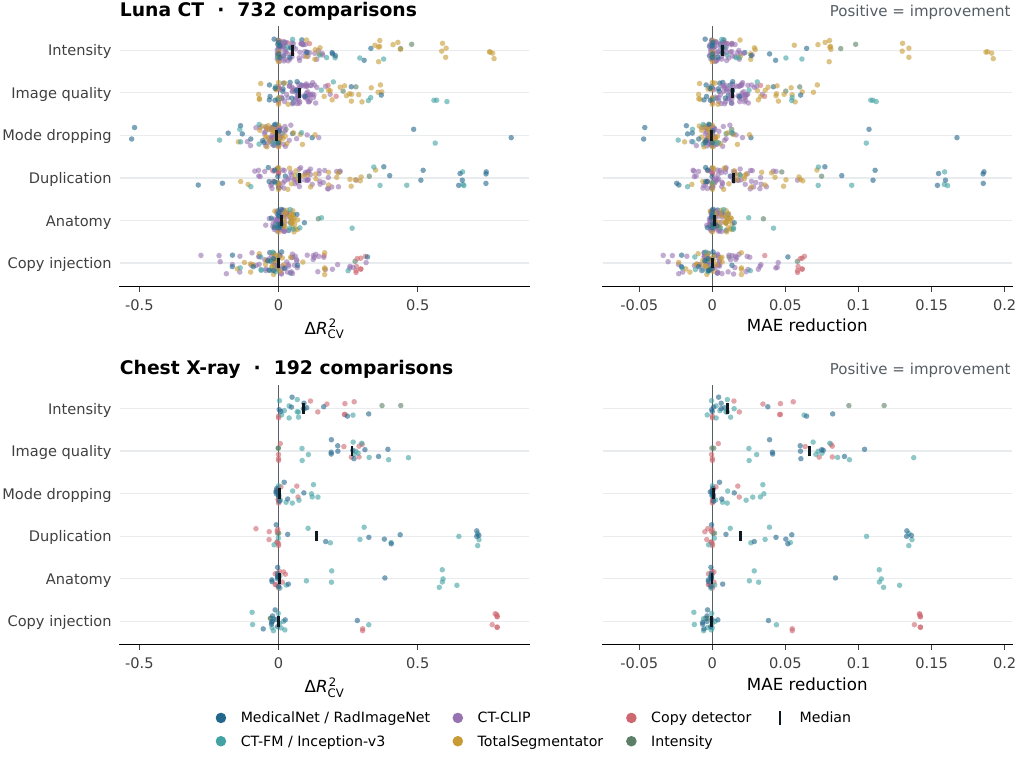}
\caption{Severity prediction after adding one metric to FID. The 924 comparisons include 732 for CT and 192 for radiography. Positive $R^2$ gains and MAE reductions favour adding a metric. Colours show measurement sources. Black ticks show family medians. Both measures use the same fixed jitter.}\label{fig:supp-complementarity-full-screen}\end{figure}

\begin{figure}[H]\centering
\includegraphics[width=\linewidth,height=0.72\textheight,keepaspectratio]{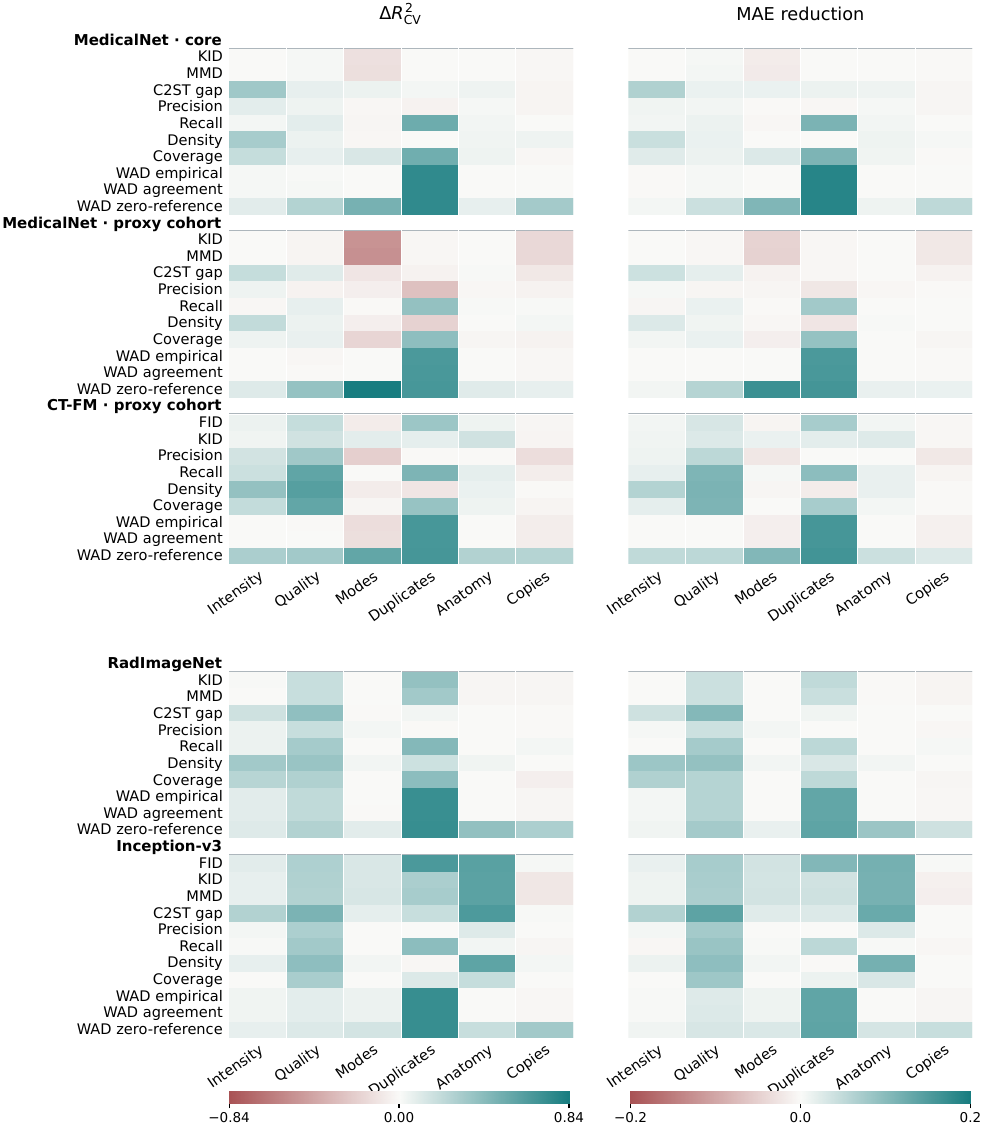}
\caption{Feature-based subset of Fig.~\ref{fig:supp-complementarity-full-screen} (174 CT and 126 radiographic comparisons). Cells show $R^2$ gain and MAE reduction after adding each metric to primary FID. Teal shows improvement, red deterioration and white little change. The scales are linear, symmetric and unsaturated. MedicalNet uses core and proxy cohorts. CT-FM uses the proxy cohort, and both radiographic extractors use the core cohort. See Table~\ref{tab:metric-overview} for WAD-DIV variants.}\label{fig:supp-complementarity-feature-variants}\end{figure}

The examples in Fig.~\ref{fig:supp-perturbation-examples} use the first source case in core replicate 0, with the original operators and stored seed. The displayed radiograph has effective PNG spacings of 0.336 mm per row and 0.410 mm per column. CT slice repetition uses the full 269-slice stack and copies each affected slice from its clean predecessor. Displacement tapers to zero at the boundary of the dashed deformation ellipse.

\begin{figure}[H]
\centering
\includegraphics[width=\linewidth,height=0.68\textheight,keepaspectratio]{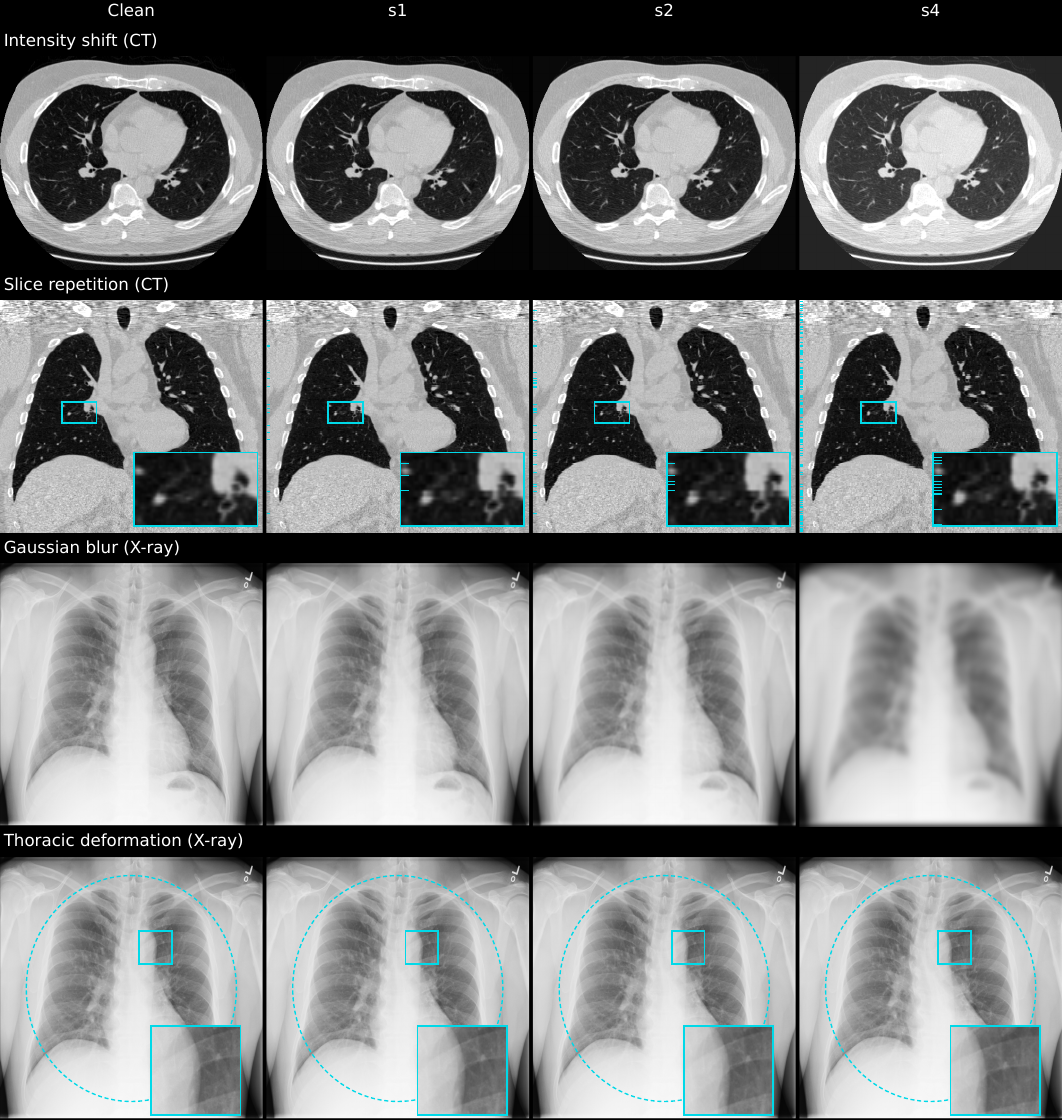}
\caption{Controlled perturbation examples. Columns show the clean image and ranks $s_1$, $s_2$ and $s_4$. The CT rows show offsets (0, 25, 50, 200 HU) and slice repetition (0, 0.05, 0.10, 0.40). The radiographic rows show blur ($\sigma=0,0.5,1,4$ mm) and deformation (0, 4, 8, 32 pixels). Cyan ticks mark 13, 27 and 108 repeated slices. Cyan boxes mark matched insets. The CT inset shows slice 140 replaced with slice 139 (one-based). Crops and display scales are fixed within each row. CT uses $[-1000,400]$ HU.}
\label{fig:supp-perturbation-examples}
\end{figure}

\FloatBarrier
\section{Autoencoder grid}
\begin{figure}[H]\centering
\includegraphics[width=\linewidth,height=0.76\textheight,keepaspectratio]{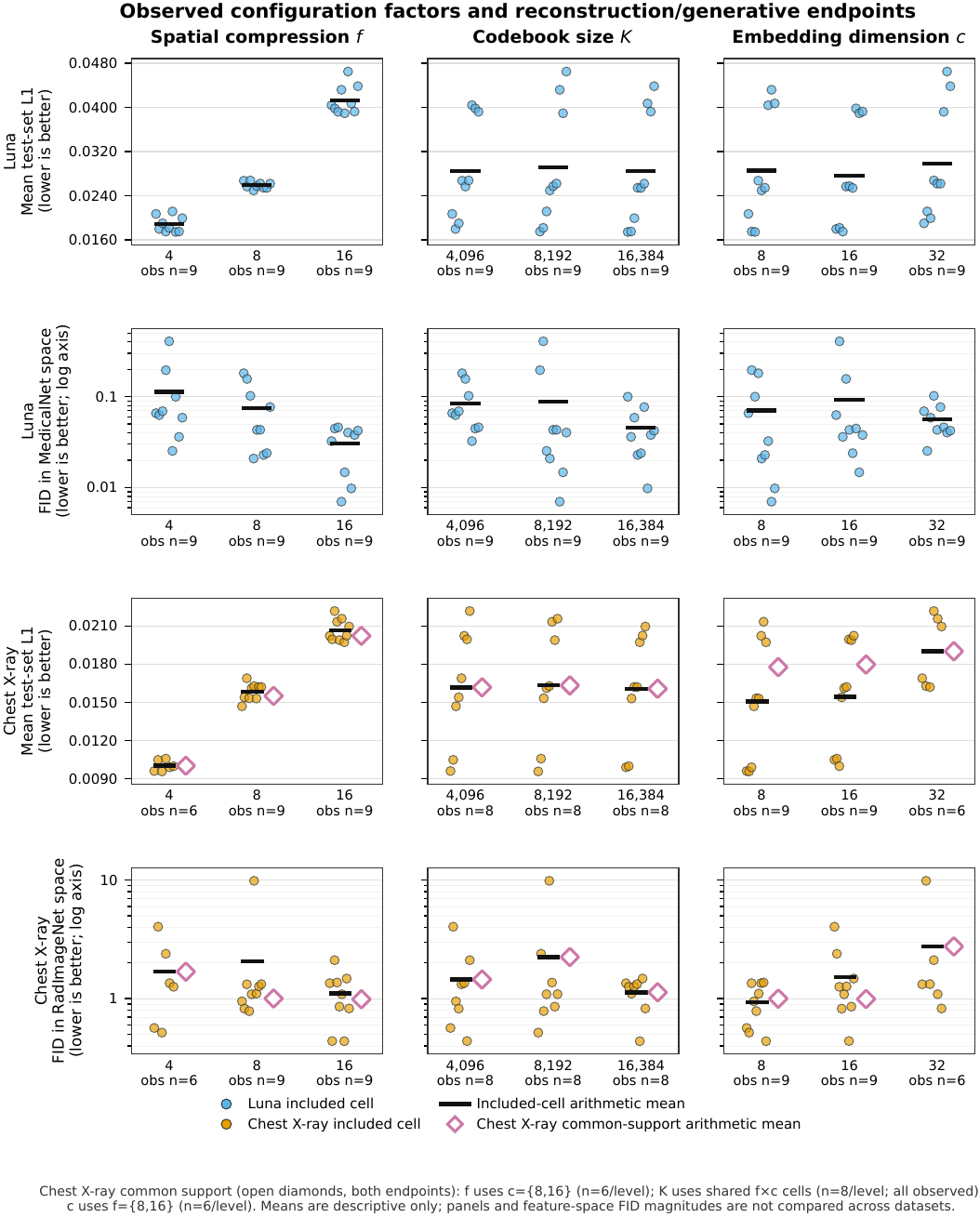}
\caption{L1 and primary FID by compression factor, codebook size and embedding dimension. Black bars show unweighted means. Open diamonds show means for the balanced radiographic subsets (Section~\ref{sec:supp-architecture}).}\label{fig:supp-factor-endpoints}\end{figure}

\begin{figure}[H]\centering
\includegraphics[width=\linewidth,height=0.76\textheight,keepaspectratio]{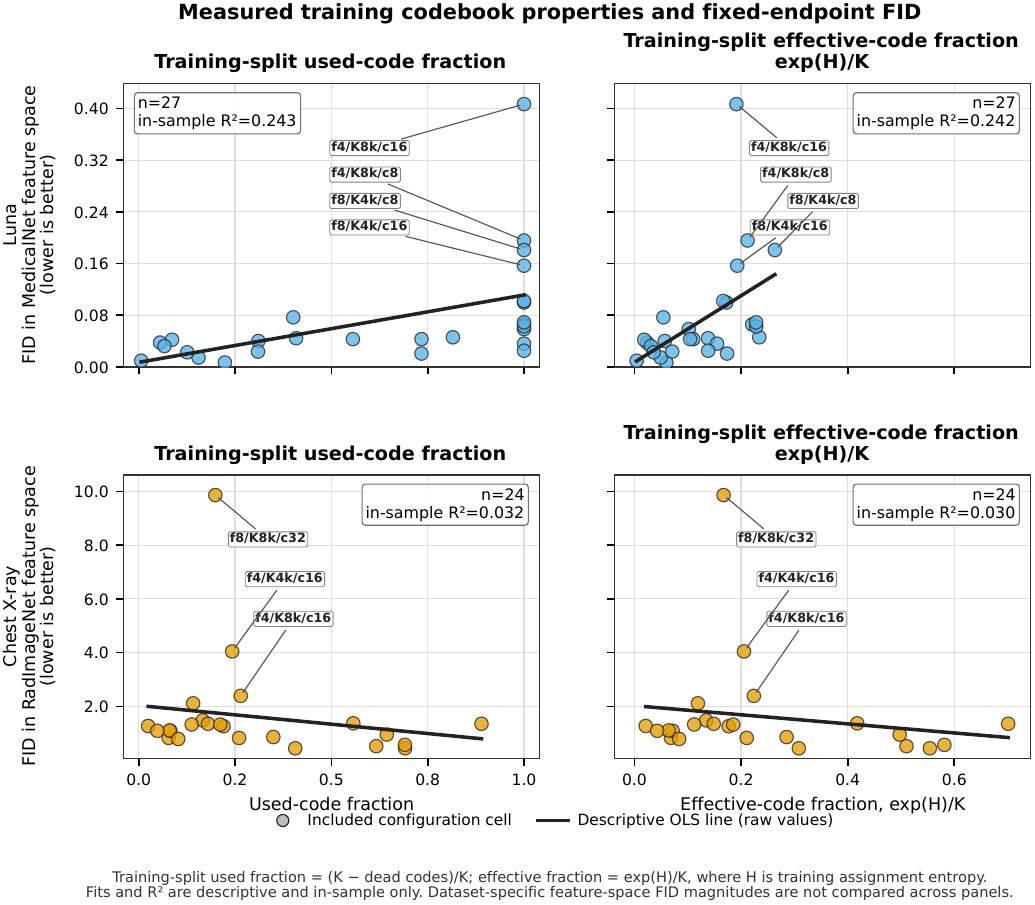}
\caption{Training codebook usage versus primary FID. The panels show the used-code and effective-code fractions. Lines are linear fits to the raw values.}\label{fig:supp-codebook-fid}\end{figure}

\begin{figure}[H]\centering
\includegraphics[width=\linewidth,height=0.39\textheight,keepaspectratio]{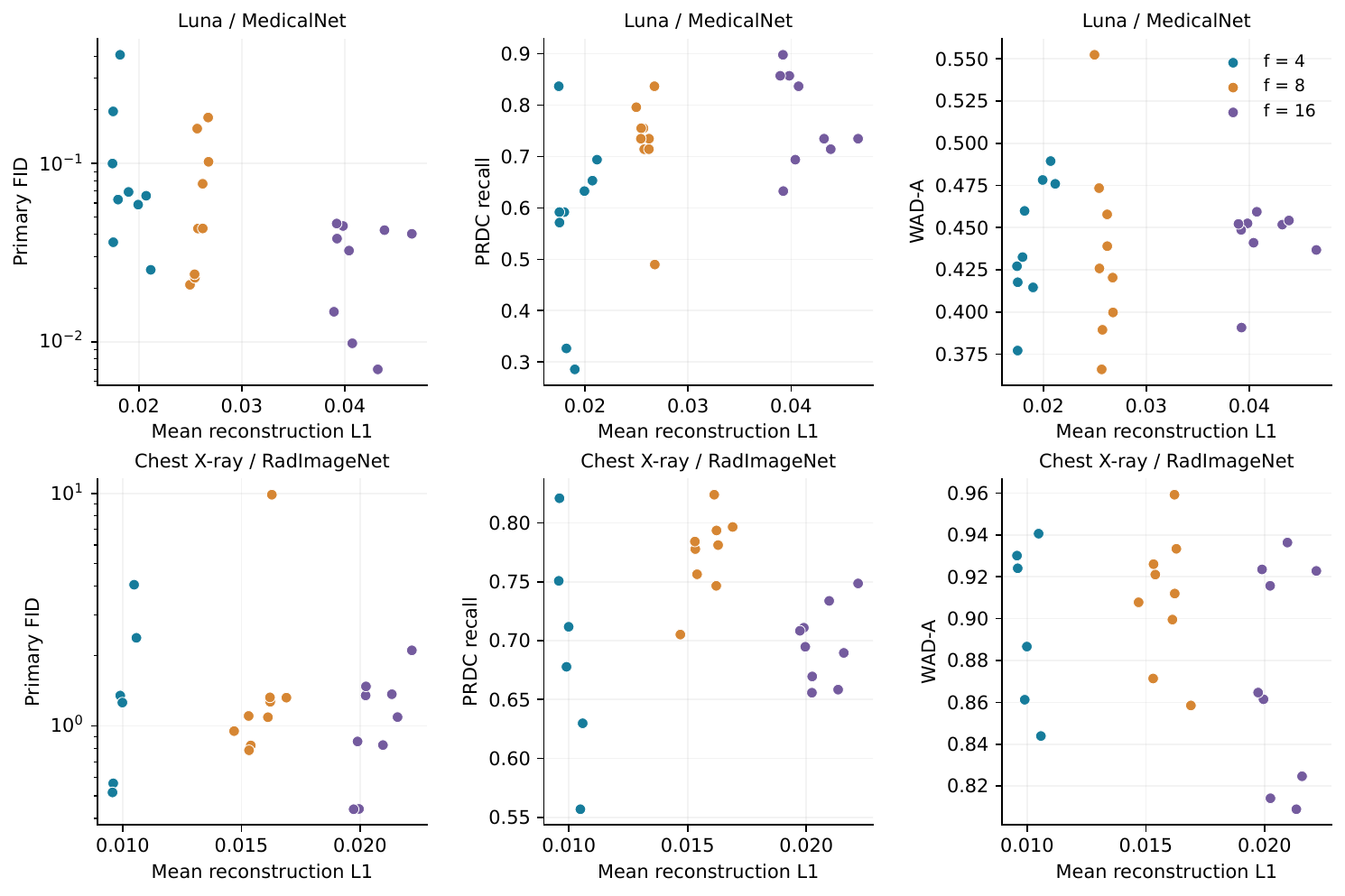}
\caption{L1 versus primary FID, PRDC recall and WAD-A. Each point represents one grid configuration. Colours show compression factor $f$.}\label{fig:supp-multimetric-l1}\end{figure}
\begin{figure}[H]\centering
\includegraphics[width=\linewidth,height=0.36\textheight,keepaspectratio]{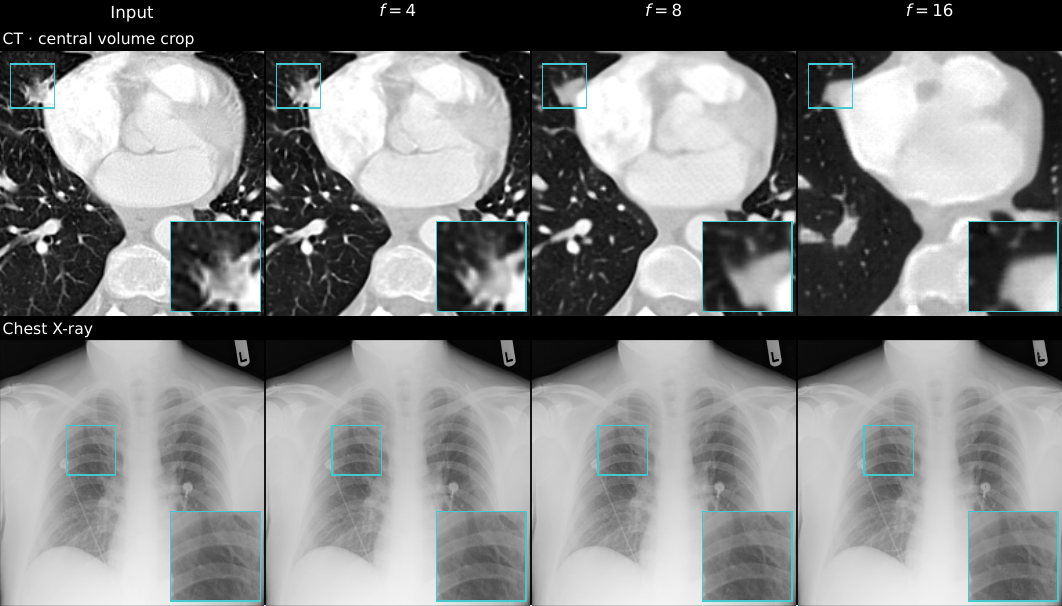}
\caption{Paired reconstructions of the first stored test case in each dataset. Insets show the same regions. CT configurations $(f,K,c)$ are $(4,16384,8)$, $(8,8192,8)$ and $(16,8192,8)$. Radiography uses $(4,8192,8)$, $(8,4096,8)$ and $(16,16384,8)$. CT shows the central axial slice from a $192\times192\times128$ crop at $[-1000,400]$ HU. Radiographs use the full $1024^2$ input after min--max normalisation, with a shared $[-1,1]$ display range. Reconstructions use the frozen checkpoints on CPU.}\label{fig:supp-paired-reconstructions}\end{figure}

\begin{figure}[H]
\centering
\includegraphics[width=\linewidth,height=0.68\textheight,keepaspectratio]{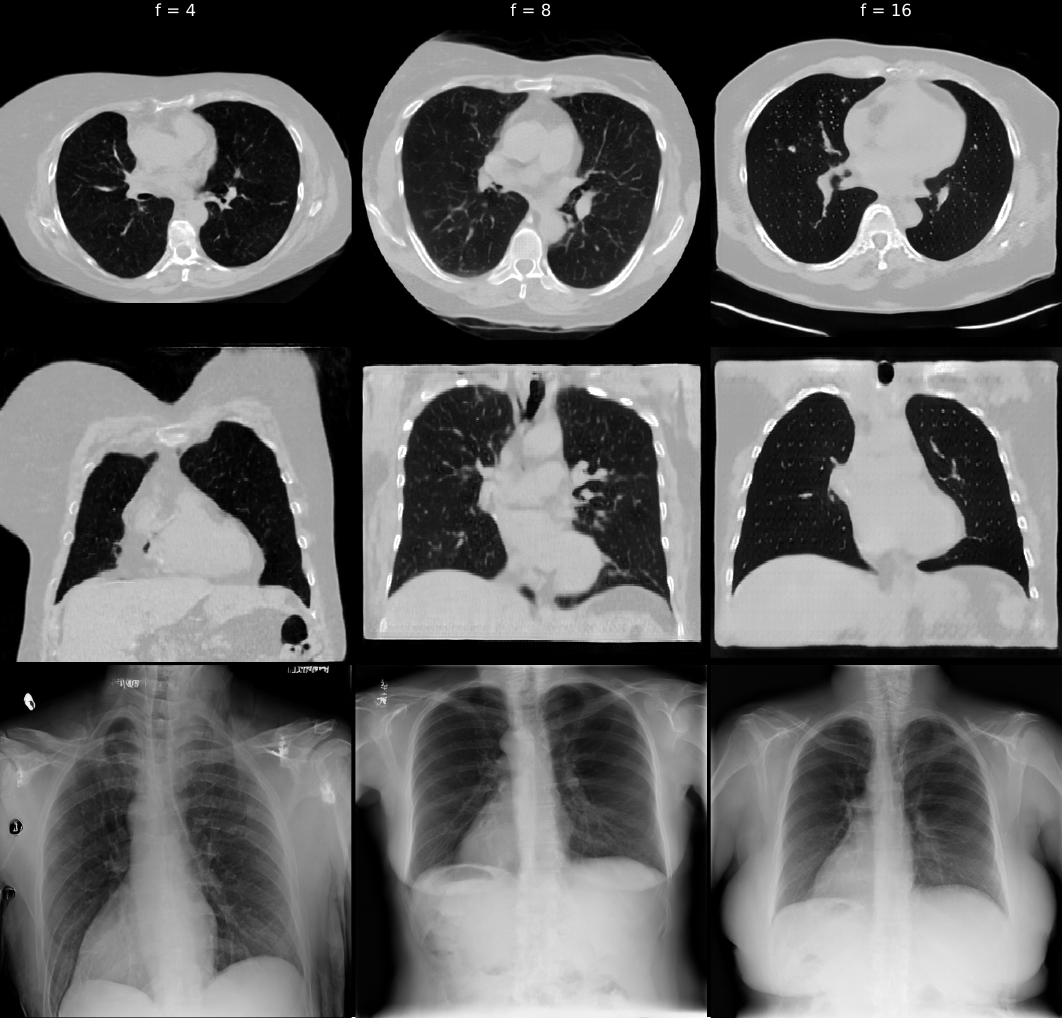}
\caption{Unpaired generations from stored sample 0 of each model in Fig.~\ref{fig:supp-paired-reconstructions}. The middle column uses the lowest-L1 model at $f=8$. CT rows show central axial and coronal slices at $[-1000,400]$ HU. Peripheral background is cropped and tiles preserve aspect ratio.}
\label{fig:supp-configuration-examples}
\end{figure}

\begin{figure}[H]
    \centering
    \includegraphics[width=\linewidth]{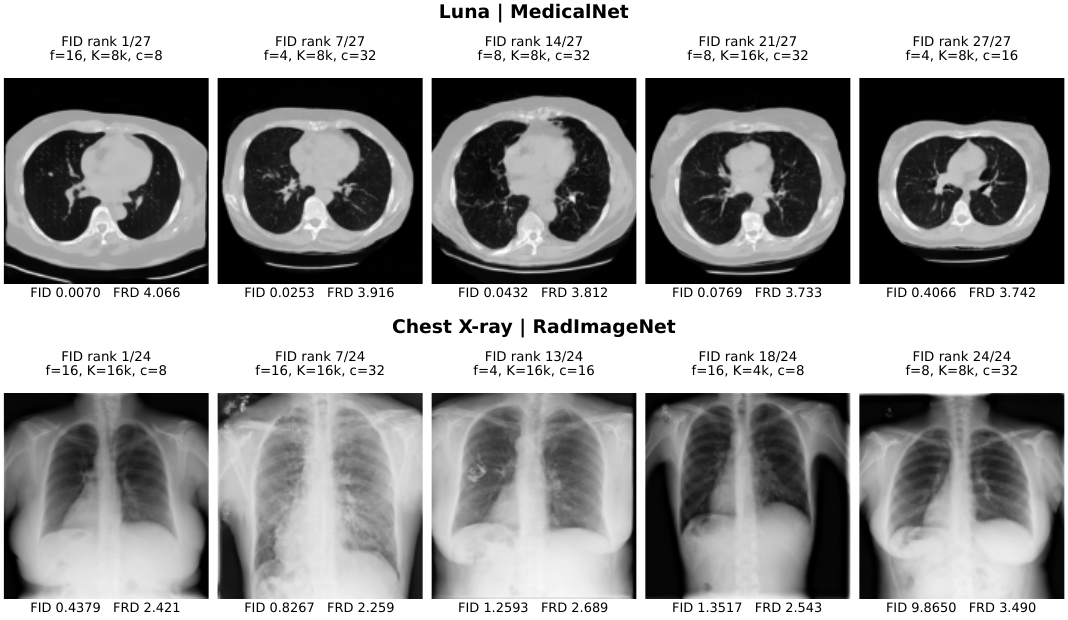}
    \caption{Unpaired samples across FID ranks. Each of five models per dataset shows its first stored sample. Labels give cohort-level FID and FRD. CT shows the central axial slice at $[-1000,400]$ HU.}
    \label{fig:metric-labelled-examples}
\end{figure}

\begin{figure}[H]
    \centering
    \includegraphics[
        width=\linewidth,
        height=0.78\textheight,
        keepaspectratio
    ]{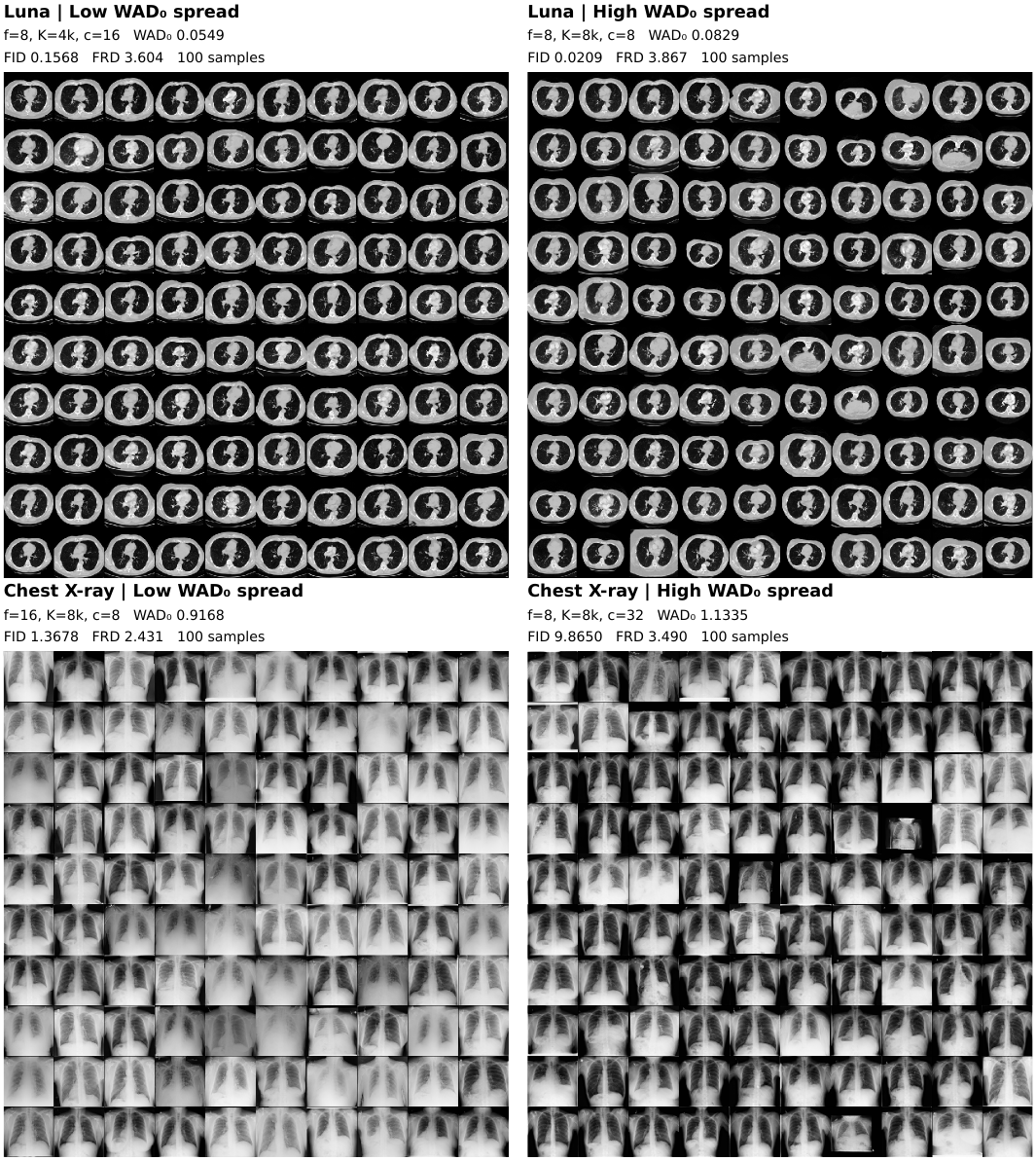}
    \caption{Samples from the models with the lowest and highest raw zero-reference WAD-DIV in each grid. Each $10\times10$ mosaic shows 100 distinct samples at roughly even intervals in stored order. CT shows central axial slices at $[-1000,400]$ HU.}
    \label{fig:wad-diversity-mosaics}
\end{figure}

\clearpage
\subsection{Complete baseline results}

\begin{table}[H]
\centering\small
\setlength{\tabcolsep}{3pt}
\caption{CT baseline grid results.}\label{tab:supp-grid-Luna}
\begin{tabular}{rrr*{11}{r}}
\toprule
$f$ & $K$ & $c$ & \shortstack{FID\\$\downarrow$} & \shortstack{C2ST\\$\downarrow$} & \shortstack{Prec.\\$\uparrow$} & \shortstack{Rec.\\$\uparrow$} & \shortstack{Den.\\$\uparrow$} & \shortstack{Cov.\\$\uparrow$} & \shortstack{WAD-A$_1$\\$\uparrow$} & \shortstack{WAD-A$_2$\\$\uparrow$} & \shortstack{FRD\\$\downarrow$} & \shortstack{$N_{\rm copy}$} & \shortstack{$N_{\rm mem}$} \\
\midrule
4 & 4096 & 8 & 0.0659 & 0.109 & 0.958 & 0.653 & 0.856 & \textbf{1} & 0.489 & \textbf{0.7} & 3.91 & 468 & 202 \\
4 & 4096 & 16 & 0.0627 & 0.155 & 0.994 & 0.592 & 1.13 & 0.939 & 0.433 & 0.638 & 3.54 & 396 & 202 \\
4 & 4096 & 32 & 0.0692 & 0.264 & 0.927 & 0.286 & 0.574 & 0.878 & 0.415 & 0.64 & 4.2 & 538 & 129 \\
4 & 8192 & 8 & 0.196 & 0.235 & \textbf{1} & 0.592 & 1.27 & 0.837 & 0.377 & 0.63 & 3.82 & 489 & 142 \\
4 & 8192 & 16 & 0.407 & 0.285 & 0.81 & 0.327 & 0.823 & 0.714 & 0.46 & 0.608 & 3.74 & 517 & 141 \\
4 & 8192 & 32 & 0.0253 & 0.115 & 0.941 & 0.694 & 1 & \textbf{1} & 0.476 & 0.661 & 3.92 & 484 & 175 \\
4 & 16384 & 8 & 0.0999 & 0.182 & 0.983 & 0.837 & 1.06 & 0.98 & 0.427 & 0.659 & 3.6 & 374 & 257 \\
4 & 16384 & 16 & 0.0361 & 0.114 & 0.986 & 0.571 & 1.25 & \textbf{1} & 0.418 & 0.664 & \textbf{3.38} & 407 & 195 \\
4 & 16384 & 32 & 0.0587 & 0.105 & 0.983 & 0.633 & 1.03 & 0.98 & 0.478 & 0.699 & 3.54 & 505 & 271 \\
8 & 4096 & 8 & 0.181 & 0.227 & 0.983 & 0.837 & 0.918 & 0.98 & 0.42 & 0.647 & 3.76 & 586 & 340 \\
8 & 4096 & 16 & 0.157 & 0.234 & \textbf{1} & 0.755 & \textbf{1.27} & 0.939 & 0.366 & 0.601 & 3.6 & 529 & 226 \\
8 & 4096 & 32 & 0.102 & 0.188 & 0.936 & 0.49 & 1.06 & 0.98 & 0.4 & 0.577 & 3.96 & 513 & 173 \\
8 & 8192 & 8 & 0.0209 & 0.0672 & 0.866 & 0.796 & 0.831 & \textbf{1} & \textbf{0.552} & 0.648 & 3.87 & 491 & 285 \\
8 & 8192 & 16 & 0.0431 & 0.128 & 0.991 & 0.714 & 1.2 & \textbf{1} & 0.389 & 0.56 & 3.65 & 494 & 241 \\
8 & 8192 & 32 & 0.0432 & 0.087 & 0.99 & 0.735 & 1.24 & \textbf{1} & 0.439 & 0.639 & 3.81 & 505 & 371 \\
8 & 16384 & 8 & 0.0228 & \textbf{0.0294} & 0.996 & 0.755 & 1.22 & \textbf{1} & 0.426 & 0.599 & 3.93 & 461 & 294 \\
8 & 16384 & 16 & 0.0239 & 0.0607 & 0.984 & 0.735 & 1.14 & \textbf{1} & 0.473 & 0.626 & 3.62 & 550 & 387 \\
8 & 16384 & 32 & 0.0769 & 0.167 & 0.924 & 0.714 & 0.821 & \textbf{1} & 0.458 & 0.637 & 3.73 & 596 & 310 \\
16 & 4096 & 8 & 0.0324 & 0.115 & 0.93 & 0.694 & 1.02 & \textbf{1} & 0.441 & 0.57 & 3.97 & 595 & 374 \\
16 & 4096 & 16 & 0.0446 & 0.126 & 0.965 & 0.857 & 1.13 & \textbf{1} & 0.452 & 0.604 & 3.74 & 598 & 422 \\
16 & 4096 & 32 & 0.046 & 0.121 & 0.99 & \textbf{0.898} & 1.15 & \textbf{1} & 0.449 & 0.626 & 3.74 & 582 & 440 \\
16 & 8192 & 8 & \textbf{0.007} & 0.0526 & 0.964 & 0.735 & 1.13 & \textbf{1} & 0.452 & 0.553 & 4.07 & 625 & 380 \\
16 & 8192 & 16 & 0.0147 & 0.0669 & 0.984 & 0.857 & 1.17 & \textbf{1} & 0.452 & 0.609 & 3.84 & 639 & 454 \\
16 & 8192 & 32 & 0.0403 & 0.121 & 0.964 & 0.735 & 0.977 & \textbf{1} & 0.437 & 0.575 & 4.14 & 487 & 301 \\
16 & 16384 & 8 & 0.0098 & 0.0815 & 0.976 & 0.837 & 1.18 & \textbf{1} & 0.459 & 0.601 & 3.95 & 603 & 444 \\
16 & 16384 & 16 & 0.0379 & 0.119 & 0.97 & 0.633 & 1.02 & 0.98 & 0.391 & 0.594 & 3.74 & 567 & 317 \\
16 & 16384 & 32 & 0.0422 & 0.12 & 0.96 & 0.714 & 1.11 & 0.98 & 0.454 & 0.613 & 4.03 & 552 & 386 \\
\bottomrule\end{tabular}
\par\smallskip\parbox{\linewidth}{\footnotesize FID, C2ST and PRDC use MedicalNet. WAD-A$_1$ uses MedicalNet and WAD-A$_2$ uses CT-FM. Bold marks the best unrounded value within each comparison, including ties. Detector counts use the training cohorts in Section~\ref{sec:supp-detector-methods}. Grid values are rounded to three significant figures. The full-precision baseline data retain both feature representations.}
\end{table}

\begin{table}[H]
\centering\small
\setlength{\tabcolsep}{3pt}
\caption{Chest X-ray baseline grid results.}\label{tab:supp-grid-ChestXray}
\begin{tabular}{rrr*{11}{r}}
\toprule
$f$ & $K$ & $c$ & \shortstack{FID\\$\downarrow$} & \shortstack{C2ST\\$\downarrow$} & \shortstack{Prec.\\$\uparrow$} & \shortstack{Rec.\\$\uparrow$} & \shortstack{Den.\\$\uparrow$} & \shortstack{Cov.\\$\uparrow$} & \shortstack{WAD-A$_1$\\$\uparrow$} & \shortstack{WAD-A$_2$\\$\uparrow$} & \shortstack{FRD\\$\downarrow$} & \shortstack{$N_{\rm copy}$} & \shortstack{$N_{\rm mem}$} \\
\midrule
4 & 4096 & 8 & 0.566 & 0.0828 & 0.856 & 0.821 & 0.556 & 0.773 & 0.924 & 0.879 & 2.26 & 426 & 364 \\
4 & 4096 & 16 & 4.05 & 0.333 & 0.888 & 0.557 & 0.464 & 0.235 & 0.941 & \textbf{0.985} & 2.8 & 313 & 224 \\
4 & 8192 & 8 & 0.517 & 0.132 & 0.91 & 0.751 & 0.665 & 0.78 & 0.93 & 0.905 & 2.69 & 369 & 338 \\
4 & 8192 & 16 & 2.39 & 0.177 & 0.9 & 0.63 & 0.612 & 0.729 & 0.844 & 0.851 & 2.72 & 435 & 344 \\
4 & 16384 & 8 & 1.35 & 0.225 & 0.916 & 0.678 & 0.672 & 0.625 & 0.861 & 0.865 & 2.75 & 343 & 306 \\
4 & 16384 & 16 & 1.26 & 0.138 & 0.888 & 0.712 & 0.635 & 0.764 & 0.887 & 0.887 & 2.69 & 327 & 323 \\
8 & 4096 & 8 & 0.95 & 0.181 & 0.952 & 0.705 & \textbf{0.857} & 0.738 & 0.908 & 0.903 & 2.87 & 402 & 375 \\
8 & 4096 & 16 & 0.824 & 0.107 & 0.927 & 0.756 & 0.736 & 0.806 & 0.921 & 0.907 & 2.59 & 418 & 420 \\
8 & 4096 & 32 & 1.32 & 0.161 & 0.899 & 0.797 & 0.618 & 0.78 & 0.858 & 0.846 & 2.54 & 442 & 402 \\
8 & 8192 & 8 & 0.785 & 0.137 & 0.9 & 0.778 & 0.631 & 0.768 & 0.926 & 0.878 & 2.48 & 374 & 368 \\
8 & 8192 & 16 & 1.09 & 0.118 & 0.876 & \textbf{0.824} & 0.612 & 0.811 & 0.9 & 0.857 & 2.33 & 432 & 445 \\
8 & 8192 & 32 & 9.86 & 0.191 & \textbf{0.959} & 0.781 & 0.806 & 0.546 & 0.933 & 0.87 & 3.49 & 397 & 319 \\
8 & 16384 & 8 & 1.1 & 0.145 & 0.916 & 0.784 & 0.714 & 0.829 & 0.871 & 0.887 & 2.57 & 422 & 420 \\
8 & 16384 & 16 & 1.27 & 0.227 & 0.916 & 0.794 & 0.68 & 0.665 & 0.912 & 0.849 & 2.45 & 404 & 394 \\
8 & 16384 & 32 & 1.33 & 0.213 & 0.916 & 0.747 & 0.71 & 0.62 & \textbf{0.959} & 0.885 & 2.76 & 248 & 236 \\
16 & 4096 & 8 & 1.35 & 0.0903 & 0.95 & 0.656 & 0.83 & 0.787 & 0.916 & 0.857 & 2.54 & 682 & 575 \\
16 & 4096 & 16 & 0.439 & 0.111 & 0.946 & 0.695 & 0.841 & \textbf{0.855} & 0.861 & 0.851 & 2.5 & 765 & 714 \\
16 & 4096 & 32 & 2.11 & 0.0714 & 0.947 & 0.749 & 0.849 & 0.83 & 0.923 & 0.805 & 2.61 & 543 & 508 \\
16 & 8192 & 8 & 1.37 & 0.161 & 0.949 & 0.658 & 0.824 & 0.784 & 0.809 & 0.846 & 2.43 & 878 & 670 \\
16 & 8192 & 16 & 0.857 & 0.0797 & 0.957 & 0.711 & 0.835 & 0.827 & 0.924 & 0.837 & 2.47 & 703 & 642 \\
16 & 8192 & 32 & 1.09 & 0.123 & 0.926 & 0.69 & 0.763 & 0.81 & 0.825 & 0.77 & 2.44 & 648 & 595 \\
16 & 16384 & 8 & \textbf{0.438} & 0.113 & 0.94 & 0.708 & 0.807 & 0.847 & 0.865 & 0.845 & 2.42 & 672 & 620 \\
16 & 16384 & 16 & 1.48 & 0.155 & 0.938 & 0.67 & 0.787 & 0.808 & 0.814 & 0.815 & 2.48 & 683 & 629 \\
16 & 16384 & 32 & 0.827 & \textbf{0.0412} & 0.914 & 0.734 & 0.725 & 0.832 & 0.936 & 0.834 & \textbf{2.26} & 591 & 499 \\
\bottomrule\end{tabular}
\par\smallskip\parbox{\linewidth}{\footnotesize FID, C2ST and PRDC use RadImageNet. WAD-A$_1$ uses RadImageNet and WAD-A$_2$ uses Inception-v3 (labelled ImageNet in some plots). See Table~\ref{tab:supp-grid-Luna} for notation.}
\end{table}

\clearpage

\FloatBarrier
\section{Complementary evaluation}
\begin{figure}[H]\centering
\includegraphics[width=\linewidth,height=0.72\textheight,keepaspectratio]{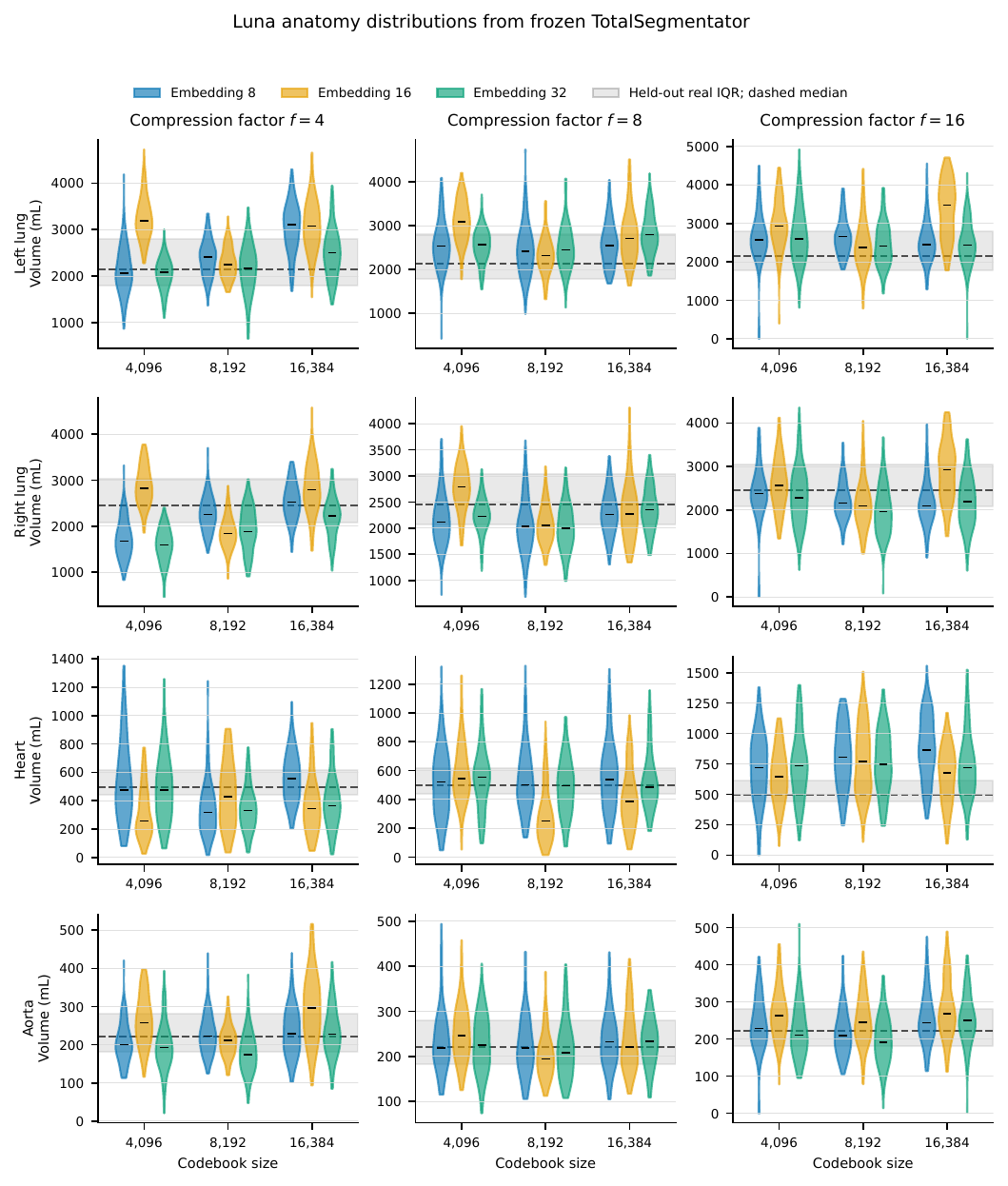}
\caption{Organ-volume distributions. At $(f,K,c)=(16,4096,8)$, zero marks one failed mask per lung, two for the heart and three for the aorta. Axes vary by organ.}\label{fig:supp_anatomy_volume}\end{figure}

\begin{figure}[H]\centering
\includegraphics[width=\linewidth,height=0.76\textheight,keepaspectratio]{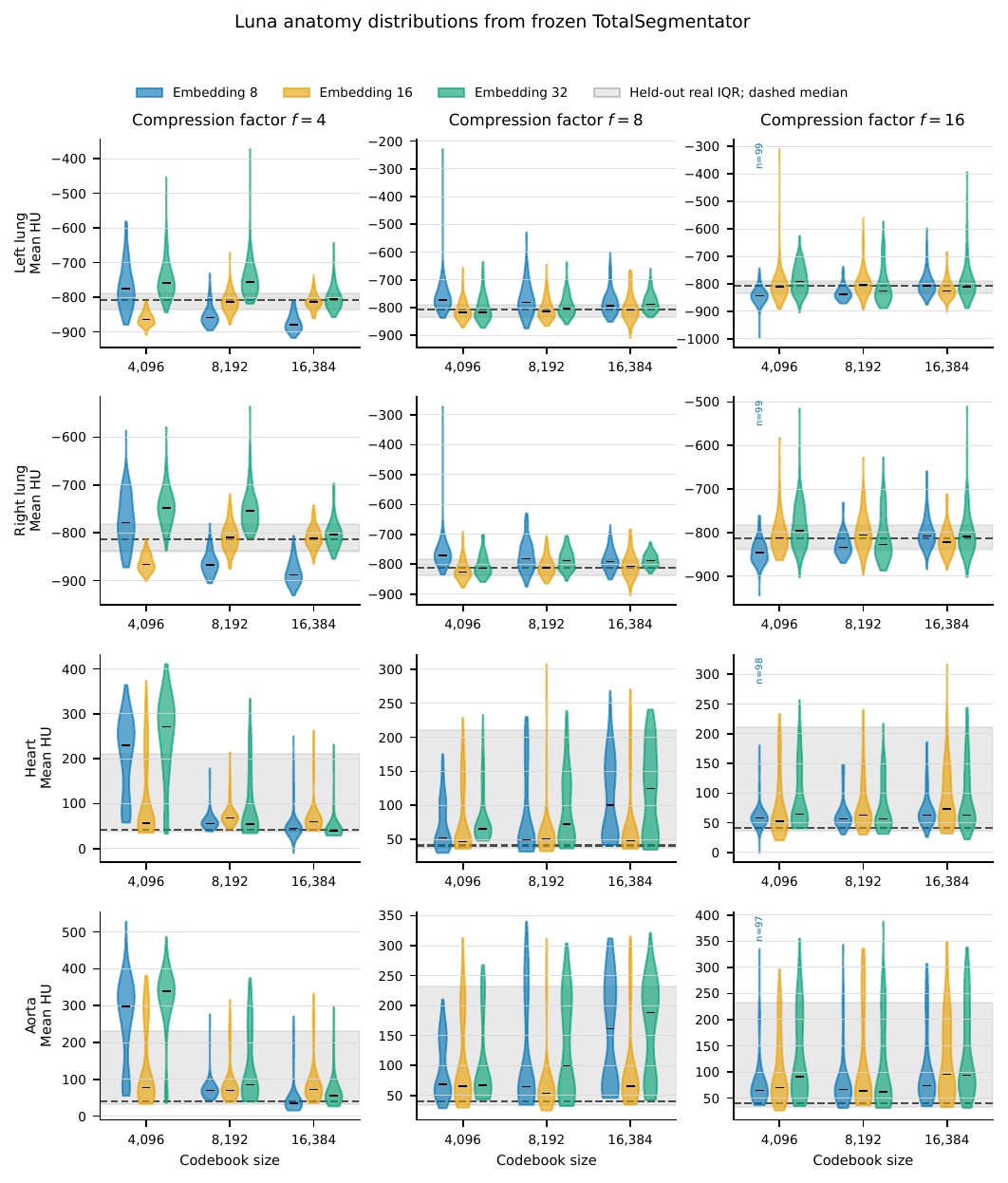}
\caption{Mean HU within each organ. At $(f,K,c)=(16,4096,8)$, the left lung, right lung, heart and aorta have 99, 99, 98 and 97 valid masks, respectively. Axes vary by organ.}\label{fig:supp_anatomy_hu}\end{figure}

\begin{figure}[H]\centering
\includegraphics[width=\linewidth,height=0.56\textheight,keepaspectratio]{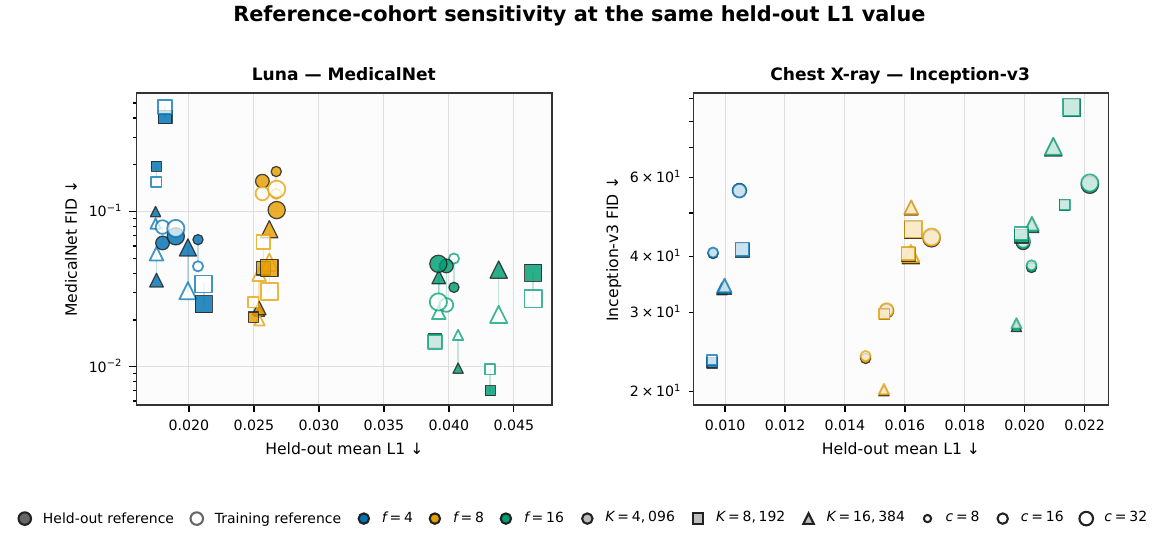}
\caption{FID against test and training references. Both FID values for each configuration are plotted at the same held-out mean L1.}\label{fig:supp-reference-cohort}\end{figure}

\FloatBarrier
\section{U-Net capacity sensitivity}
\begin{table}[H]
\centering
\small
\setlength{\tabcolsep}{4pt}
\caption{U-Net capacity results at each compression factor.}\label{tab:supp-capacity-results}
\begin{tabular}{lrlrrrrrrrr}
\toprule
Dataset & $f$ & U-Net & \shortstack{FID\\$\downarrow$} & \shortstack{Prec.\\$\uparrow$} & \shortstack{Rec.\\$\uparrow$} & \shortstack{Den.\\$\uparrow$} & \shortstack{Cov.\\$\uparrow$} & \shortstack{WAD-A\\$\uparrow$} & \shortstack{$N_{\rm copy}$} & \shortstack{$N_{\rm mem}$} \\
\midrule
Luna & 4 & Smaller & 0.3713 & 0.532 & \textbf{0.878} & 0.358 & 0.796 & \textbf{0.581} & 450 & 96 \\
Luna & 4 & Baseline & 0.0999 & 0.983 & 0.837 & 1.058 & 0.980 & 0.427 & 374 & 257 \\
Luna & 4 & Wider & 0.0278 & 0.988 & 0.510 & \textbf{1.260} & 0.980 & 0.412 & 486 & 284 \\
Luna & 8 & Smaller & 0.0203 & 0.946 & 0.633 & 0.964 & \textbf{1.000} & 0.448 & 558 & 278 \\
Luna & 8 & Baseline & 0.0209 & 0.866 & 0.796 & 0.831 & \textbf{1.000} & 0.552 & 491 & 285 \\
Luna & 8 & Wider & 0.0311 & 0.968 & 0.673 & 1.032 & \textbf{1.000} & 0.420 & 593 & 222 \\
Luna & 16 & Smaller & 0.0123 & \textbf{0.999} & 0.592 & 1.189 & \textbf{1.000} & 0.381 & 556 & 395 \\
Luna & 16 & Baseline & 0.0147 & 0.984 & 0.857 & 1.174 & \textbf{1.000} & 0.452 & 639 & 454 \\
Luna & 16 & Wider & \textbf{0.0057} & 0.973 & 0.857 & 1.111 & \textbf{1.000} & 0.435 & 682 & 473 \\
Chest X-ray & 4 & Smaller & 1.0932 & 0.921 & 0.737 & 0.714 & 0.685 & 0.898 & 252 & 245 \\
Chest X-ray & 4 & Baseline & 0.5174 & 0.910 & 0.751 & 0.665 & 0.780 & 0.930 & 369 & 338 \\
Chest X-ray & 4 & Wider & 1.3199 & 0.912 & 0.779 & 0.616 & 0.594 & 0.874 & 388 & 365 \\
Chest X-ray & 8 & Smaller & 1.2512 & 0.921 & 0.739 & 0.701 & 0.723 & \textbf{0.959} & 309 & 310 \\
Chest X-ray & 8 & Baseline & 0.9495 & \textbf{0.952} & 0.705 & \textbf{0.857} & 0.738 & 0.908 & 402 & 375 \\
Chest X-ray & 8 & Wider & 2.3848 & 0.891 & \textbf{0.797} & 0.607 & 0.735 & 0.879 & 559 & 521 \\
Chest X-ray & 16 & Smaller & 0.9173 & 0.922 & 0.741 & 0.750 & 0.810 & 0.870 & 409 & 390 \\
Chest X-ray & 16 & Baseline & 0.4379 & 0.940 & 0.708 & 0.807 & \textbf{0.847} & 0.865 & 672 & 620 \\
Chest X-ray & 16 & Wider & \textbf{0.4147} & 0.929 & 0.689 & 0.752 & 0.844 & 0.897 & 1068 & 921 \\
\bottomrule
\end{tabular}
\par\smallskip\parbox{\linewidth}{\footnotesize Feature-based scores use MedicalNet for CT and RadImageNet for chest X-ray. Section~\ref{sec:supp-architecture} gives the fixed autoencoders and training settings. Table~\ref{tab:supp-grid-Luna} gives the shared conventions.}
\end{table}

\begin{figure}[H]\centering
\includegraphics[width=\linewidth,height=0.37\textheight,keepaspectratio]{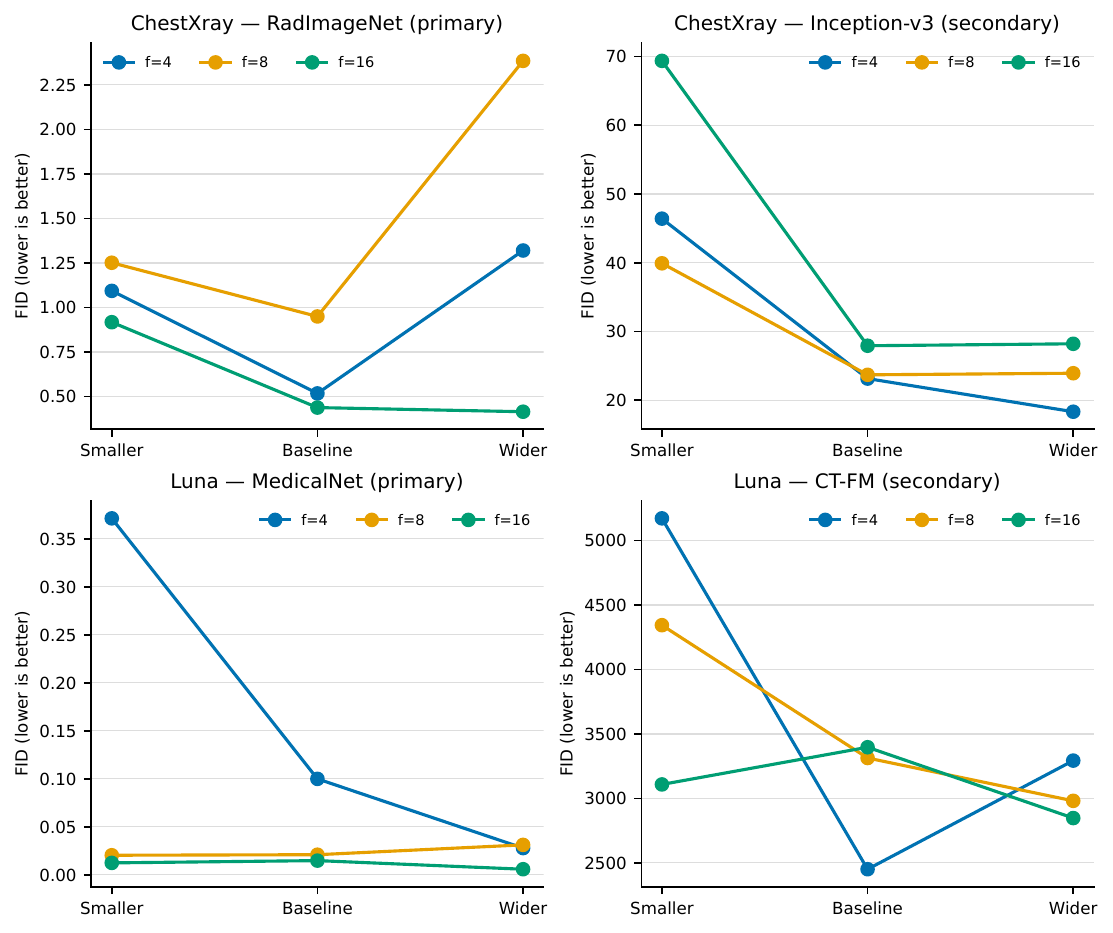}
\caption{FID by U-Net width and feature extractor. Compare values within each panel. Section~\ref{sec:supp-architecture} gives the CT training settings.}\label{fig:supp-capacity-fid}\end{figure}

\FloatBarrier
\section{Generator-family comparison}
\begin{table}[H]
\centering
\small
\setlength{\tabcolsep}{4pt}
\caption{Generator comparison by dataset.}\label{tab:supp-generators}
\begin{tabular}{llrrrrrrrrrr}
\toprule
Dataset & Generator & $f$ & \shortstack{FID\\$\downarrow$} & \shortstack{Rank\\$\downarrow$} & \shortstack{Prec.\\$\uparrow$} & \shortstack{Rec.\\$\uparrow$} & \shortstack{Den.\\$\uparrow$} & \shortstack{Cov.\\$\uparrow$} & \shortstack{WAD-A\\$\uparrow$} & \shortstack{$N_{\rm copy}$} & \shortstack{$N_{\rm mem}$} \\
\midrule
Luna & U-Net LDM & 4 & 0.0999 & 7/8 & 0.983 & 0.837 & 1.058 & 0.980 & 0.427 & 374 & 257 \\
Luna & DiT & 4 & 0.0213 & 5/8 & \textbf{0.991} & 0.878 & \textbf{1.267} & \textbf{1.000} & 0.469 & 518 & 403 \\
Luna & U-Net LDM & 8 & 0.0209 & 4/8 & 0.866 & 0.796 & 0.831 & \textbf{1.000} & \textbf{0.552} & 491 & 285 \\
Luna & DiT & 8 & \textbf{0.0061} & \textbf{1/8} & 0.976 & \textbf{0.898} & 1.176 & \textbf{1.000} & 0.511 & 499 & 446 \\
Luna & U-Net LDM & 16 & 0.0147 & 3/8 & 0.984 & 0.857 & 1.174 & \textbf{1.000} & 0.452 & 639 & 454 \\
Luna & DiT & 16 & 0.0113 & 2/8 & 0.970 & 0.735 & 1.179 & \textbf{1.000} & 0.489 & 447 & 394 \\
Luna & HA-GAN & -- & 0.0425 & 6/8 & 0.959 & 0.735 & 1.016 & \textbf{1.000} & 0.470 & 790 & 5 \\
Luna & WDM-3D & -- & 0.2640 & 8/8 & 0.580 & 0.102 & 0.258 & 0.082 & 0.484 & 731 & 6 \\
Chest X-ray & U-Net LDM & 4 & 0.5174 & 5/8 & 0.910 & 0.751 & 0.665 & 0.780 & 0.930 & 369 & 338 \\
Chest X-ray & DiT & 4 & 0.4878 & 4/8 & 0.933 & 0.808 & 0.787 & 0.861 & 0.918 & 373 & 386 \\
Chest X-ray & U-Net LDM & 8 & 0.9495 & 7/8 & \textbf{0.952} & 0.705 & \textbf{0.857} & 0.738 & 0.908 & 402 & 375 \\
Chest X-ray & DiT & 8 & 0.4645 & 3/8 & 0.934 & 0.803 & 0.773 & 0.853 & 0.914 & 398 & 405 \\
Chest X-ray & U-Net LDM & 16 & 0.4379 & 2/8 & 0.940 & 0.708 & 0.807 & 0.847 & 0.865 & 672 & 620 \\
Chest X-ray & DiT & 16 & 0.5393 & 6/8 & 0.926 & 0.717 & 0.761 & 0.826 & 0.891 & 470 & 471 \\
Chest X-ray & HA-GAN-2D & -- & 1.7393 & 8/8 & 0.794 & 0.374 & 0.478 & 0.401 & 0.834 & 280 & 141 \\
Chest X-ray & WaveDiff & -- & \textbf{0.2495} & \textbf{1/8} & 0.927 & \textbf{0.880} & 0.828 & \textbf{0.939} & \textbf{0.975} & 326 & 341 \\
\bottomrule
\end{tabular}
\par\smallskip\parbox{\linewidth}{\footnotesize Ranks use unrounded primary FID. A dash marks a model with no VQ-GAN compression factor. See Table~\ref{tab:supp-grid-Luna} for shared conventions.}
\end{table}

\begin{figure}[H]
\centering
\includegraphics[width=\linewidth,height=0.40\textheight,keepaspectratio]{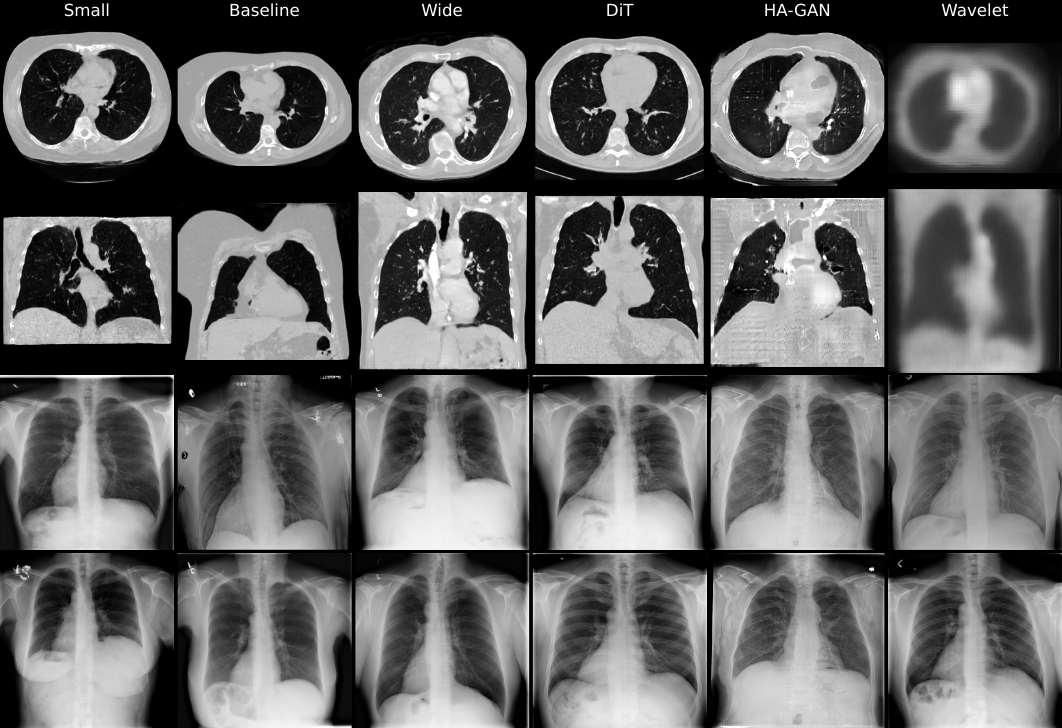}
\caption{Unpaired samples by generator. Columns show three U-Net widths, DiT, HA-GAN and wavelet diffusion (WDM-3D for CT, WaveDiff for chest X-ray). Latent models use the $f=4$ autoencoder. CT rows show central axial and coronal slices of sample 0 at $[-1000,400]$ HU with background cropped. Matched coronal views use the training voxel aspect ratio. The $256^3$ exports keep their native ratio. Radiographic rows show samples 0 and 1.}
\label{fig:supp-architecture-examples}
\end{figure}

\FloatBarrier
\section{Training proximity and closest matches}\label{sec:supp-memorisation}
Tables~\ref{tab:supp-grid-Luna}--\ref{tab:supp-generators} cover 38 CT and 35 radiographic cohorts. They include 27 CT and 24 radiographic baselines, plus six extra capacity models, three DiTs and two contextual models per dataset. Baseline entries shared across tables use the same cohorts.

\begin{figure}[H]
\centering
\includegraphics[width=\linewidth,height=0.68\textheight,keepaspectratio]{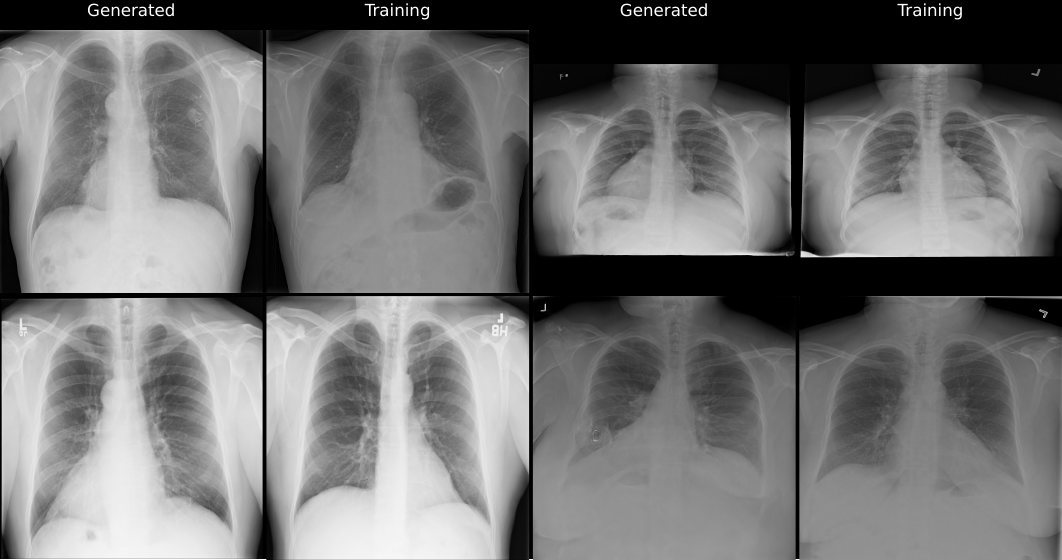}
\caption{Closest matches to training radiographs. In row order, the L1-selected U-Net, FID-selected U-Net, $f=4$ DiT and WaveDiff pairs have correlations of 0.6224, 0.7608, 0.5849 and 0.6535. Stored grayscale, orientation and aspect ratio are preserved. Dark padding is cropped.}
\label{fig:supp-closest-matches}
\end{figure}

\begin{figure}[htbp]
\centering
\includegraphics[width=\linewidth,height=0.68\textheight,keepaspectratio]{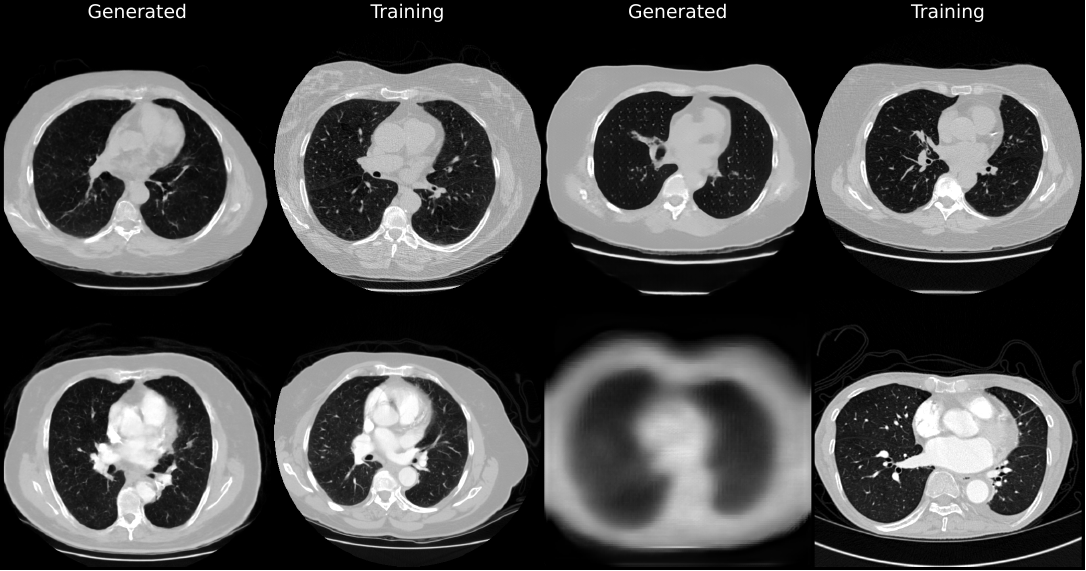}
\caption{Closest matches to training CT volumes. In row order, the L1-selected U-Net $(4,16384,8)$, FID-selected U-Net $(16,8192,8)$, $f=4$ DiT and WDM-3D pairs have correlations of 0.9820, 0.9976, 0.9870 and 0.9916. Central axial slices use $[-1000,400]$ HU and preserve orientation and aspect ratio.}
\label{fig:supp-luna-closest-matches}
\end{figure}

\clearpage
\endgroup